\documentclass[12pt,journal,draftcls]{IEEEtran}
\ifCLASSINFOpdf
\else
\fi
\usepackage{pstricks,graphicx}
\usepackage{amssymb}
\usepackage{amsmath}
\usepackage{setspace}
\usepackage{psfrag}

\newtheorem{theorem}{Theorem}[section]
\newtheorem{lemma}[theorem]{Lemma}

\newtheorem{proposition}[theorem]{Proposition}

\newtheorem{remark}[theorem]{Remark}

\newenvironment{proof}[1][Proof]{\begin{trivlist}
\item[\hskip \labelsep {\bfseries #1}]}{\end{trivlist}}

\newcommand{\qed}{\nobreak \ifvmode \relax \else
      \ifdim\lastskip<1.5em \hskip-\lastskip
      \hskip1.5em plus0em minus0.5em \fi \nobreak
      \vrule height0.75em width0.5em depth0.25em\fi}

\begin{document}
%
\title{\Huge Bounding the Rate Region of Vector Gaussian Multiple Descriptions with Individual and Central Receivers}

%
%

\author{Guoqiang~Zhang,
        W.~Bastiaan~Kleijn,~\IEEEmembership{Fellow,~IEEE,}
        and~Jan~{\O}stergaard,~\IEEEmembership{Member,~IEEE}
\thanks{The work was partly presented at the Data Compression Conference, 2010.}
\thanks{G.~Zhang (guoqiang.zhang@ee.kth.se) and W.~B.~Kleijn (bastiaan.kleijn@ee.kth.se) are with the school
of Electrical Engineering, KTH-Royal Institute of Technology.}
\thanks{J.~{\O}stergaard (janoe@ieee.org) is with the Department of Electronic Systems, Aalborg University, Aalborg, Denmark.}
}

\onecolumn
\doublespacing
\maketitle

\begin{abstract}

In this work, the rate region of the vector Gaussian multiple
description problem with individual and central quadratic
distortion constraints is studied. In particular, an outer bound
to the rate region of the L-description problem is derived. The
bound is obtained by lower bounding a weighted sum rate for each
supporting hyperplane of the rate region. The key idea is to
introduce at most L-1 auxiliary random variables and further
impose upon the variables a Markov structure according to the
ordering of the description weights. This makes it possible to
greatly simplify the derivation of the outer bound. In the scalar
Gaussian case, the complete rate region is fully characterized by
showing that the outer bound is tight. In this case, the optimal
weighted sum rate for each supporting hyperplane is obtained by
solving a single maximization problem. This contrasts with
existing results, which require solving a min-max optimization
problem.

\end{abstract}


\begin{IEEEkeywords}
Multiple description coding, rate region, entropy power
inequality.
\end{IEEEkeywords}

%
\IEEEpeerreviewmaketitle

\section{Introduction}

%

%
%
%
\IEEEPARstart{M}{ultiple} description (MD) coding is a joint
source-channel coding scheme aimed at combating unreliable
communication links. The basic principle is to generate a set of
descriptions for the information source with the property that any
subset of the descriptions provides an approximation of the source
with a certain fidelity. A natural research goal is to determine
the transmission limits under some quality of service requirement.
Generally, there are $2^L-1$ distortion constraints for
the $L$-description case, each corresponding to a particular
combination of received descriptions. Thus, the problem complexity
increases exponentially with an increasing number of descriptions.

The pioneering work on the multiple description problem is the two-description
achievable rate region by El Gamal and Cover
\cite{Gamal82Multiple} (EGC scheme). The region was shown to be
tight for a memoryless Gaussian source by Ozarow
\cite{Ozarow80Multiple}, and was shown not to be tight in general
when there is no excess rate (no redundancy) by Zhang and Berger
\cite{Zhang87MDC}. This result was further extended to the
$L$-description case in \cite{Venkataramani03MDC_bound}. Later,
the work in \cite{Pradhan04Multiple},\cite{Ruri05MDCLowerbound},
provided an enlarged achievable rate region by using the random
binning ideas from distributed source coding.

The characterization of the (tight) rate region for the general
$L$-description problem is quite challenging, and remains an open
problem. Instead, the main focus has been on special cases, e.g.,
the case of symmetric side distortion constraints
\cite{TianMDCRateRegionApp}, or the case where only a subset of
the distortion constraints is of concern
\cite{Wang07VectorMDC,Wang09MDCGaussian,Chen09MDCRateRegion}.

For the particular case that only individual (only one description
is received) and central receivers are of importance, Wang and
Viswanath \cite{Wang07VectorMDC} derived the optimal sum rate for
the vector Gaussian source. In \cite{Chen09MDCRateRegion}, Chen
considered the same transmission scenario, and derived the rate
region. However, his work was limited to the scalar Gaussian
source. The direct extension of his work to the vector Gaussian
source appears to be difficult if possible.

The present work considers the rate region of the vector Gaussian
multiple description problem with individual and central distortion
constraints. One major contribution is that an outer bound is
derived for the considered rate region. The outer bound is
formulated by lower-bounding a weighted sum rate associated with a
supporting hyper-plane of the rate region. The expression for the
bound only involves a maximization process.

Another contribution is that when the new approach is applied to
the scalar Gaussian multiple description problem, the derived
outer bound is shown to be tight, thus fully characterizing the
rate region. Note that the expression for the bound corresponds to
a maximization problem as compared to that in
\cite{Chen09MDCRateRegion} which involves a min-max game (or a
max-min game), thus significantly reducing the problem complexity.

We now summarize the notations used in this paper. Lower case
letters denote scalar random variables and boldface lower case
letters denote vector random variables. Boldface upper case
letters are used to denote matrices. Specifically, we use
$\boldsymbol{0}$ and $\boldsymbol{I}$ to denote the all-zero
matrices (including zero vectors) and the identity matrices,
respectively. We also use $\boldsymbol{H}$ to denote all-one
matrices. The operator $|\cdot|$ refers to the determinant of a
matrix, unless otherwise specified. The operator
$\mathbb{E}(\cdot)$ denotes the expectation. For random vectors
$\boldsymbol{y}_1$ and $\boldsymbol{y}_2$, we use
$\mathbb{E}\left[\boldsymbol{y}_1|\boldsymbol{y}_2\right]$ to
denote the conditional expectation of $\boldsymbol{y}_1$ given
$\boldsymbol{y}_2$, and
$\textrm{Cov}[\boldsymbol{y}_1|\boldsymbol{y}_2]$ to denote the
covariance matrix of
$\boldsymbol{y}_1-\mathbb{E}\left[\boldsymbol{y}_1|\boldsymbol{y}_2\right]$.
The partial order $\succ$ ($\succeq$) denotes positive definite
(semidefinite) ordering: $\boldsymbol{A}\succ\boldsymbol{B}$
$(\boldsymbol{A}\succeq\boldsymbol{B})$ means that
$\boldsymbol{A}-\boldsymbol{B}$ is a positive definite
(semidefinite) matrix. All the logarithm functions are to base
$e$.

\section{Problem Formulation and Main Results}
In this section we first define the multiple description rate-region problem
formally. We then introduce the Gaussian multiple description
scheme for the problem, which provides an achievable rate region.
Finally we present the main results of the paper.
\subsection{Problem Formulation}
\label{subsec:problem_formuation} Suppose the information source
is an i.i.d. process $\{\boldsymbol{x}[m]\}$ with marginal
distribution $\mathcal{N}(\boldsymbol{0},\boldsymbol{K}_x)$, i.e.,
a collection of i.i.d. real Gaussian random vectors. Let the
covariance matrix $\boldsymbol{K}_x$ be an $N\times N$ positive
definite matrix. There are $L$ encoding functions at the
transmitter, each encoding a source sequence of length $n$,
$\boldsymbol{x}^n=\left(\boldsymbol{x}[1]^t,\ldots,\boldsymbol{x}[n]^t\right)^t$.
Denote the resulting codewords as $f^{(n)}_l(x^n)$,
$l=1,\ldots,L$. Let $C_l^{(n)}$, $l=1,\ldots,L$, denote the
corresponding codebooks, i.e., $f^{(n)}_l(x^n)\in C_l^{(n)}$. The
output of the $l$'th encoder is sent through the $l$'th
communication channel at rate $R_l=\frac{1}{n}\log|C_l^{(n)}|$,
where $|C_l^{(n)}|$ denotes the cardinality of the codebook.

In response to the $L$ descriptions, there are $L$ individual
receivers and one central receiver. The $l$'th individual receiver
uses its received codeword to generate an approximation $g_l(
f^{(n)}_l(x^n))$ to the source sequence $\boldsymbol{x}^n$,
$l=1,\ldots,L$. On the other hand, the central receiver generates
an approximation to the source sequence based on all $L$
codewords. Since we are only interested in the quadratic
distortion measure, the optimal approximation is given by the
minimal mean-squared error (MMSE) estimation of the source
sequence. We say a rate vector $(R_1,\ldots,R_L)$ is achievable if
there exist, for all sufficiently large $n$, encoders of rates
$(R_1,\ldots,R_L)$ and decoders, such that
\begin{small}\begin{equation}
\hspace{2mm}\frac{1}{n}\sum_{m=1}^n \textrm{Cov}\left[\boldsymbol{x}[m]|f_l^{(n)}(\boldsymbol{x}^n)\right]\preceq \boldsymbol{D}_l, \quad l=1,\ldots,L, \\
\vspace{-3mm}\end{equation}
\begin{equation}
\frac{1}{n}\sum_{m=1}^n
\textrm{Cov}\left[\boldsymbol{x}[m]|f_l^{(n)}(\boldsymbol{x}^n),\ldots,f_L^{(n)}(\boldsymbol{x}^n)\right]\preceq
\boldsymbol{D}_0.
\end{equation}\end{small}\hspace{-1mm}
The rate region
$\mathcal{R}(\boldsymbol{K}_x,\boldsymbol{D}_1,\ldots,\boldsymbol{D}_L,\boldsymbol{D}_0)$
is the convex hull of the set of all achievable rate vectors
subject to the individual side distortion constraints
$\boldsymbol{D}_l$, $l=1,\ldots,L$, and the central distortion
constraint $\boldsymbol{D}_0$. Throughout the paper, we consider
$\boldsymbol{0}\prec \boldsymbol{D}_0\prec \boldsymbol{D}_l$,
$\forall l \in \{1,\ldots,L\}$.

We focus on characterizing the rate region
$\mathcal{R}(\boldsymbol{K}_x,\boldsymbol{D}_1,\ldots,\boldsymbol{D}_L,\boldsymbol{D}_0)$.
Our work is an extension of \cite{Wang07VectorMDC} which only
addressed the optimal sum-rate problem for the above transmission
scenario. There are two motivations behind our work. First
obtaining the rate region is of great interest from a theoretic
point of view. Second the rate-region expression can provide
insight in designing efficient asymmetric (i.e, the rates are
different across the channels) multiple description systems.

Motivated by the fact that
$\mathcal{R}(\boldsymbol{K}_x,\boldsymbol{D}_1,\ldots,\boldsymbol{D}_L,\boldsymbol{D}_0)$
is a closed convex set, we consider characterizing its supporting
hyperplane, which can be formulated as an optimization problem of
the form:
\begin{small}\begin{equation}
\min_{(R_1,\ldots,R_L)\in \mathcal{R}(\boldsymbol{K}_x,\boldsymbol{D}_1,\ldots,\boldsymbol{D}_L,\boldsymbol{D}_0)} \sum_{l=1}^L\beta_l R_l,
\label{equ:weighted_sume_optimization}
\end{equation}\end{small}\hspace{-1mm}
where $\beta_l$, $l=1,\ldots,L$, are arbitrary positive
parameters. The case where at least one of the weighting factors
is zero defines a trivial problem. To see this, notice that one
can always put enough rate to the descriptions associated with the
zero weighting factors, in order to meet the central distortion
constraint. Then one simply needs to use the minimal
single-description rate for each remaining description in order to
satisfy the individual distortion constraints and thereby achieve
the minimal (optimal) weighted sum rate.

Without loss of generality, we may assume $\beta_1\geq
\ldots\geq\beta_L>0$, which can always be obtained by rearranging
the description indices. It may happen that some weighting factors
are equal. We further group the weighting factors
$(\beta_2,\beta_2,\beta_3,\ldots,\beta_L)$ (the first element is
modified on purpose, and it will be explained later) according to
their values. The elements within each group have the same value
and different groups take different values. Denote the resulting
weighting factors as  $\alpha_1>\alpha_2>\ldots>\alpha_J$, where
$\alpha_j$ is associated with $m_j$ rates in
(\ref{equ:weighted_sume_optimization}). With this, it follows that
$\sum_{j=1}^Jm_j=L$, and $m_1\geq 2$. The maximum achievable value
of $J$ is $L-1$. Denote
\begin{small}\begin{equation}M_1^j=\sum_{i=1}^j m_i\quad j=1,\ldots,J\end{equation}\end{small}\hspace{-1mm}
and let $M_1^0=0$. The optimization problem
(\ref{equ:weighted_sume_optimization}) can be rewritten as
\begin{small}\begin{equation}
\min_{(R_1,\ldots,R_L)\in \mathcal{R}(\boldsymbol{K}_x,\boldsymbol{D}_1,\ldots,\boldsymbol{D}_L,\boldsymbol{D}_0)} \alpha_0 R_1+\sum_{j=1}^J\alpha_j\sum_{i=1}^{m_j}\left(R_{M_1^{j-1}+i}\right),
\label{equ:weighted_sume_optimization2}
\end{equation}\end{small}\hspace{-1mm}
where $\alpha_0=\beta_1-\beta_2$. The reason to distinguish
different weighting factors is to facilitate the derivation of the
outer bound to (\ref{equ:weighted_sume_optimization}). It will be
shown later that the choice of $\alpha_0$ is of little importance.
It doest not affect the construction of the outer bound.

\subsection{Gaussian Description Scheme}
\label{subsec:Gaussian_description} In this subsection, we
introduce the Gaussian description scheme, with which an
achievable rate region is then obtained. Conceptually, in the
transmission of a vector Gaussian source $\boldsymbol{x}$, $L$
parallel noisy sub-channels with additive Gaussian noises are
constructed. This has the advantage that the transmission
distortions and rates can be easily analyzed. Let
$\boldsymbol{w}_1$, $\ldots$, $\boldsymbol{w}_L$ be
$N$-dimensional Gaussian vectors independent of $\boldsymbol{x}$,
of which the marginal distributions are denoted as
$\mathcal{N}(\boldsymbol{0},\boldsymbol{K}_l)$, $l=1,\ldots,L$.
Define
\begin{small}\begin{equation}
\boldsymbol{u}_l=\boldsymbol{x}+\boldsymbol{w}_l, \quad l=1,\ldots,L.
\end{equation}\end{small}\hspace{-1mm}
Such a channel is referred to as a \emph{Gaussian test channel}
\cite{zamir99GaussianOuterBound}. For any subset
$S\subseteq\{1,\ldots,L\}$, we use $\boldsymbol{K}_{S}$ to denote
the covariance matrix of all $\boldsymbol{w}_l$, $l\in S$. With a
slight abuse of notation, we refer to
$\boldsymbol{K}_{{\{1,\ldots,L\}}}$ as $\boldsymbol{K}_w$.

We consider the MMSE estimation of $\boldsymbol{x}$ using subsets
of $\boldsymbol{u}_l$, $l=1,\ldots,L$. It can be shown that
\begin{small}\begin{equation}
\textrm{Cov}[\boldsymbol{x}|\boldsymbol{u}_l, l\in S]=\left(\boldsymbol{K}_x^{-1}
+(\boldsymbol{I}_N,\ldots,\boldsymbol{I}_N)\boldsymbol{K}_{S}^{-1}(\boldsymbol{I}_N,\ldots,\boldsymbol{I}_N)^t\right)^{-1}, \quad \forall S\subseteq \{1,
\ldots,L\}.
\label{equ:MMSE_x}
\end{equation}\end{small}\hspace{-1mm}
Define
\begin{small}\begin{equation}
\boldsymbol{\underline{K}}_S\stackrel{\Delta}{=}\left[(\boldsymbol{I}_N,\ldots,\boldsymbol{I}_N)\boldsymbol{K}_{S}^{-1}(\boldsymbol{I}_N,\ldots,\boldsymbol{I}_N)^t\right]^{-1},
\quad S\subseteq \{1,\ldots,L\}.
\label{equ:virtual_channel_matrix}
\end{equation}\end{small}\hspace{-1mm}
It is immediate that
$\boldsymbol{\underline{K}}_l=\boldsymbol{K}_l$, $l=1,\ldots,L$.
Similarly to the introduction of $\boldsymbol{K}_w$, we let
$\boldsymbol{\underline{K}}_w=\boldsymbol{\underline{K}}_{\{1,\ldots,L\}}$.
To clarify, for each $S\subseteq\{1,\ldots,L\}$ with cardinality
$|S|$, $\boldsymbol{K}_S$ is of size $N|S|\times N|S|$, whereas
$\underline{\boldsymbol{K}}_S$ is of size $N\times N$. Plugging
(\ref{equ:virtual_channel_matrix}) into (\ref{equ:MMSE_x})
produces
\begin{small}\begin{equation}
\textrm{Cov}[\boldsymbol{x}|\boldsymbol{u}_l, l\in S]=\left(\boldsymbol{K}_x^{-1}+\boldsymbol{\underline{K}}_S^{-1}\right)^{-1}, \quad S\subseteq \{1,\ldots,L\}.
\label{equ:MMSE_x2}
\end{equation}\end{small}\hspace{-1mm}
We can define a virtual additive Gaussian noisy channel for each
subset $S$ in (\ref{equ:MMSE_x2}), where the covariance matrix of
the Gaussian noise takes the form $\boldsymbol{\underline{K}}_S$.
Therefore, the quadratic distortion of the MMSE estimation of the
source $\boldsymbol{x}$ from this channel output gives the same
expression as (\ref{equ:MMSE_x2}).

The achievable rate vector by using the Gaussian description
scheme has been studied in detail in
\cite{Venkataramani03MDC_bound} (for the scalar source case), and
in \cite{Wang07VectorMDC} (for the vector source case). We
summarize the result in the following lemma.
\begin{lemma}[\cite{Wang07VectorMDC}]  For every $\boldsymbol{K}_w $ such that
\begin{small}\begin{eqnarray}
&&\textrm{Cov}[\boldsymbol{x}|\boldsymbol{u}_l]\preceq \boldsymbol{D}_l, \quad l=1,\ldots,L\nonumber \\
&&\textrm{Cov}[\boldsymbol{x}|\boldsymbol{u}_1,\ldots,\boldsymbol{u}_L] \preceq\boldsymbol{D}_0,
\label{equ:Gaussian_dis_constraint}
\end{eqnarray}\end{small}\hspace{-1mm}
the rate vector satisfying
\begin{small}\begin{equation}
\sum_{l\in S}R_l \geq \left[\sum_{l\in S}h(\boldsymbol{u}_l)\right]-h(\boldsymbol{u}_l,l\in S|\boldsymbol{x})
=\frac{1}{2}\log\frac{\prod_{l\in S}|\boldsymbol{K}_x+\boldsymbol{K}_l|}{|\boldsymbol{K}_{S}|}, \quad \forall S\subseteq \{1,\ldots,L\},
\label{equ:Gaussian_rate}
\end{equation}\end{small}\hspace{-1mm}
is achievable. \label{lemma:rate_region_inner}
\end{lemma}

Lemma \ref{lemma:rate_region_inner} defines an achievable rate
region. In Section \ref{section:scalar_Gaussian}, we will show
that the achievable region is tight for a scalar Gaussian source.

\subsection{Outer Bound to the Rate Region}
In this subsection we present our new outer bound to the rate
region
$\mathcal{R}(\boldsymbol{K}_x,\boldsymbol{D}_1,\ldots,\boldsymbol{D}_L,\boldsymbol{D}_0)$.
The expression of the outer bound is obtained by lower-bounding a
weighted sum rate (\ref{equ:weighted_sume_optimization2}) of every
supporting hyper-plane of the rate region.

In our construction of the outer bound in our work, we introduce a
set of auxiliary Gaussian random vectors as in
\cite{Chen09MDCRateRegion}. Differently from
\cite{Chen09MDCRateRegion}, we impose a Markov structure on those
auxiliary random vectors according to the order of the weighting
factors. The main reason to group the weighting factors
$(\beta_2,\beta_2,\beta_3,\cdots,\beta_L)$ (see Subsection
\ref{subsec:problem_formuation}) by their values is to identify
the number of auxiliary Gaussian random vectors needed for the
construction of the outer bound. For the optimization problem in
(\ref{equ:weighted_sume_optimization2}), we introduce $J$
auxiliary Gaussian random vectors, $\boldsymbol{z}_j$,
$j=1,\ldots,J$, one for each distinct weighting factor (i.e.,
$\boldsymbol{z}_j$ is associated with $\alpha_j$.). When all the
weighting factors are identical (i.e, $J=1$), only one auxiliary
random vector is needed. This special case is for deriving an
outer bound to the sum-rate, which was addressed in
\cite{Wang07VectorMDC}.

We now explain the Markov structure of the $J$ random vectors in
more detail. The covariance matrices of the auxiliary vectors are
chosen such that a high-indexed vector has a large covariance
matrix. Denote the covariance matrix of $\boldsymbol{z}_j$ as
$\boldsymbol{N}_j$, $j=1,\ldots,J$. The Markov structure can be
mathematically formulated as $\boldsymbol{0}\prec \alpha_1
\boldsymbol{N}_1\prec \alpha_2 \boldsymbol{N}_2\prec \ldots\prec
\alpha_J \boldsymbol{N}_J$. By using this partial ordering
structure, the derivation of the outer bound is significantly
simplified. Further, it facilitates the recognition of optimality
conditions for the outer bound to be tight.  On the other hand,
the work in \cite{Chen09MDCRateRegion} uses $L$ auxiliary random
variables without imposing a Markov structure on them.

Considering the multiple description coding with respect to individual and central
distortion constraints for a Gaussian source with distribution
$\mathcal{N}(\boldsymbol{0},\boldsymbol{K}_x)$, we have the
following outer bound to any supporting hyperplane of the rate
region.

\begin{theorem}
Given the distortion constraints
$(\boldsymbol{D}_1,\ldots,\boldsymbol{D}_L,\boldsymbol{D}_0)$ and
a zero-mean vector Gaussian source with covariance matrix
$\boldsymbol{K}_x$, the weighted sum rate
(\ref{equ:weighted_sume_optimization2}) is lower-bounded by
\begin{small}\begin{eqnarray}
&&\alpha_0R_1+\sum_{j=1}^{J}\alpha_j \left(\sum_{i=1}^{m_j}R_{M_1^{j-1}+i}\right) \nonumber \\
&& \geq \sup_{B(\{\boldsymbol{N}_i\}_{i=1}^J)} \Bigg( \frac{\alpha_0}{2}\log\frac{|\boldsymbol{K}_x|}{|\boldsymbol{D}_1|}
+\frac{\alpha_1}{2}\log\frac{\alpha_1^N|\boldsymbol{K}_x||\boldsymbol{K}_x+\boldsymbol{N}_1|^{m_1-1}|\boldsymbol{N}_2-\boldsymbol{N}_1|}
{(\alpha_1-\alpha_2)^N\prod_{i=1}^{m_1}|\boldsymbol{N}_1+\boldsymbol{D}_i||\boldsymbol{N}_2|} \nonumber\\
&& + \sum_{j=2}^{J-1}\frac{\alpha_j}{2}\log\frac{(\alpha_{j-1}-\alpha_j)^N|\boldsymbol{K}_x+\boldsymbol{N}_j|^{m_j}|\boldsymbol{N}_{j-1}||\boldsymbol{N}_{j+1}-\boldsymbol{N}_j|}
{(\alpha_j-\alpha_{j+1})^N\prod_{i=1}^{m_j}|\boldsymbol{N}_j+\boldsymbol{D}_{M_1^{j-1}+i}||\boldsymbol{N}_j-\boldsymbol{N}_{j-1}||\boldsymbol{N}_{j+1}|}\nonumber \\
&& + \frac{\alpha_J}{2}\log\frac{(\alpha_{J-1}-\alpha_J)^N|\boldsymbol{N}_J+\boldsymbol{D}_0||\boldsymbol{K}_x+\boldsymbol{N}_J|^{m_J}|\boldsymbol{N}_{J-1}|}
{\alpha_J^N|\boldsymbol{D}_0|\prod_{i=1}^{m_J}|\boldsymbol{N}_J+\boldsymbol{D}_{M_1^{J-1}+i}||\boldsymbol{N}_J-\boldsymbol{N}_{J-1}|} \Bigg),
\label{equ:lower_bound}
\end{eqnarray}\end{small}\hspace{-1mm}
where
$B(\{\boldsymbol{N}_i\}_{i=1}^J)=\{(\boldsymbol{N}_1,\ldots,\boldsymbol{N}_J)\in
\mathbb{R}^{N\times NJ} |\boldsymbol{0}\prec \alpha_1
\boldsymbol{N}_1\prec \alpha_2 \boldsymbol{N}_2\prec \ldots\prec
\alpha_J \boldsymbol{N}_J\}$, and $J>1$. For the special case that
$J=1$, the lower bound is expressed as
\begin{small}\begin{equation}
\alpha_0R_1+\alpha_1\sum_{l=1}^LR_l\geq \sup_{\boldsymbol{N}_1\succ\boldsymbol{0}}
\frac{\alpha_0}{2}\log\frac{|\boldsymbol{K}_x|}{|\boldsymbol{D}_1|}
+\frac{\alpha_1}{2}
\log\left(\frac{|\boldsymbol{K}_x||\boldsymbol{K}_x+\boldsymbol{N}_1|^{L-1}|\boldsymbol{D}_0+\boldsymbol{N}_1|}
{|\boldsymbol{D}_0|\prod_{l=1}^L|\boldsymbol{D}_l+\boldsymbol{N}_1|}\right).
\label{equ:lower_bound_sumRate}
\end{equation}\end{small}\hspace{-1mm}
\label{theorem:outer_bound}
\end{theorem}
\begin{proof}
See Section \ref{section:outer_bound}.
\end{proof}

One special case of $B(\{\boldsymbol{N}_i\}_{i=1}^J)$ in
(\ref{equ:lower_bound}) is of particular interest. Letting
 $\boldsymbol{N}_1=[\epsilon/\alpha_1]\boldsymbol{I}_N$,
 $\boldsymbol{N}_j=[\epsilon(1+\epsilon+\ldots,+\epsilon^{j-1})/\alpha_j]\boldsymbol{I}_N$,
 $j=2,\ldots,J$, and $\epsilon\rightarrow 0^{+}$, we
 obtain from (\ref{equ:lower_bound}) the following lower
 bound
\begin{small}\begin{equation}
\alpha_0R_1+\sum_{j=1}^{J}\alpha_j \left(\sum_{i=1}^{m_j}R_{M_1^{j-1}+i}\right)\geq \frac{\alpha_0}{2}\log\frac{|\boldsymbol{K}_x|}{|\boldsymbol{D}_1|}
+\sum_{j=1}^J\frac{\alpha_j}{2}\left(\sum_{i=1}^{m_j}\log\frac{|\boldsymbol{K}_x|}{|\boldsymbol{D}_{M_1^{j-1}+i}|}\right)\mbox{.}\label{equ:lower_bound_loose_central}
\end{equation}\end{small}\hspace{-1mm}
This bound is actually the weighted summation of the $L$ minimum
single-description rates,
$\frac{1}{2}\log\frac{|\boldsymbol{K}_x|}{|\boldsymbol{D}_i|}$,
$i=1,\ldots,L$. Note that the central distortion constraint is not
involved in the bound. For the extreme case that the central
distortion constraint is loose,
(\ref{equ:lower_bound_loose_central}) becomes a tight bound. In
this situation, any hyperplane attains the rate region at the
corner point
$\left(\frac{1}{2}\log\frac{|\boldsymbol{K}_x|}{|\boldsymbol{D}_1|},\ldots,\frac{1}{2}\log\frac{|\boldsymbol{K}_x|}{|\boldsymbol{D}_L|}\right)$.

\subsection{Optimal Weighted Sum Rate}
\label{subsec:optimal_weighted_sumrate} In this subsection, we
first present the optimality conditions for the Gaussian
description scheme to produce the optimal weighted sum rate.
Specifically, we derive the optimality conditions on the
covariance matrix $\boldsymbol{K}_w$ as introduced in subsection
\ref{subsec:Gaussian_description}. Then we discuss how to compute
the optimal matrix $\boldsymbol{K}_w$ if it exists.

For the application of the Gaussian description scheme, we provide
optimality conditions under which the performance of the scheme
reaches the outer bound (\ref{equ:lower_bound}). Therefore, if a
Gaussian description scheme can be constructed for a weighting
vector $(\alpha_0,\alpha_1,\ldots,\alpha_J)$ such that the
optimality conditions are satisfied, the corresponding optimal
weighted sum rate can be obtained.

\begin{theorem}
If, for any given $\boldsymbol{K}_x$, there exists a
$\boldsymbol{K}_w$ such that:
 \begin{enumerate}
    \item Layered correlation:
        \begin{small}\begin{equation}
        \mathbb{E}\left[\boldsymbol{w}_{M_1^{j-1}+i}\boldsymbol{w}_{k}^t\right]=-\boldsymbol{A}_j,
        \quad \forall \textrm{ } 1\leq k<M_1^{j-1}+i  \textrm{, }
        i=1,\ldots,m_j\textrm{, }j=1,\ldots,J
        \label{equ:layered_correlation}
        \end{equation}\end{small}\hspace{-1mm}
    \item Proportionality:
\begin{small}\begin{equation}
\alpha_j\left(\boldsymbol{A}_j+\boldsymbol{\underline{K}}_{\{1,\ldots,M_1^j\}}\right)^{-1}-(\alpha_j-\alpha_{j+1})\boldsymbol{\underline{K}}_{\{1,\ldots,M_1^j\}}^{-1}
=\alpha_{j+1}\left(\boldsymbol{A}_{j+1}+\boldsymbol{\underline{K}}_{\{1,\ldots,M_1^j\}}\right)^{-1}, \quad j=1,\ldots,J-1,
\label{equ:proportionality}
\end{equation}\end{small}\hspace{-1mm}
\end{enumerate}
where
$\boldsymbol{0}\prec\alpha_1\boldsymbol{A}_1\prec\alpha_2\boldsymbol{A}_2\prec\ldots\prec\alpha_J\boldsymbol{A}_J\prec
\alpha_J\boldsymbol{K}_x$, and such that the set of distortion
constraints
$(\boldsymbol{D}_1,\dotsc,\boldsymbol{D}_L,\boldsymbol{D}_0)$ is
achieved with equality, then the outer bound given by
(\ref{equ:lower_bound}) is tight.
\label{theorem:optimal_weight_sumRate}
\end{theorem}

\begin{proof}
See Section \ref{section:Opt_weight_sumRate}.
\end{proof}

The property of layered correlation in Theorem
\ref{theorem:optimal_weight_sumRate} refers to the fact that the
correlation matrix of $\boldsymbol{w}_j$ and any lower indexed
random vector $\boldsymbol{w}_k$ $k<j$ remains the same. The
number of different correlation matrices is determined by the
number of different weighting factors. For the optimization
problem (\ref{equ:weighted_sume_optimization2}), $J$ different
correlation matrices have to be constructed. Informally, each
correlation matrix $\boldsymbol{A}_j$ in
(\ref{equ:layered_correlation}) controls the redundancy among the
$m_j$ descriptions associated with $\alpha_j$, and the set of
descriptions associated with $\{\alpha_i,i<j\}$. Further, a small
correlation matrix corresponds to a high redundancy among the
associated descriptions. For the extreme case that the central
distortion constraint is loose, the optimal correlation matrices
take the form
$\boldsymbol{A}_1=\cdots=\boldsymbol{A}_J=\boldsymbol{0}$. The $L$
descriptions are highly correlated, implying a high redundancy
embedded in the descriptions. The optimal weighted sum rate for
this special case is given by
(\ref{equ:lower_bound_loose_central}).

The proportionality condition (\ref{equ:proportionality}) imposes
an ordered structure on the correlation matrices, i.e.,
$\boldsymbol{0}\prec\alpha_1\boldsymbol{A}_1\prec\alpha_2\boldsymbol{A}_2\prec\ldots\prec\alpha_J\boldsymbol{A}_J\prec
\alpha_J\boldsymbol{K}_x$. This ordered structure is essentially
determined by the ordered structure on the weighting factors,
i.e., $\alpha_1>\alpha_2>\cdots>\alpha_J$. We now explain this
property in an informal way. If the central distortion constraint
is active, each description has to carry an extra rate besides the
minimum single-description rate to account for the central
distortion constraint. Since a large weighting factor enforces a
large penalty on the associated description rate in
(\ref{equ:weighted_sume_optimization2}), it is desirable that the
descriptions with a large weighting factor carry a small amount of
extra rates. By relating a small correlation matrix with a large
weighting factor (i.e., $\boldsymbol{A}_j$ corresponds to
$\alpha_j$, $j=1,\ldots,J$), the redundancy introduced by the
correlation matrix is large, rendering a small amount of extra
rate for the descriptions associated with the large weighting
factor. (\ref{equ:proportionality}) fully characterizes the
relationship between the correlation matrices and the weighting
factors.

Next we consider how to obtain the the covariance matrix
$\boldsymbol{K}_w$ satisfying the conditions in Theorem
\ref{theorem:optimal_weight_sumRate} if it exists. We first study
the property of the layered correlation
(\ref{equ:layered_correlation}). We find that if the covariance
matrix $\boldsymbol{K}_w$ has the layered structure
(\ref{equ:layered_correlation}), the matrices
$\underline{\boldsymbol{K}}_{\{1,\ldots,M_1^j\}}$ and
$\boldsymbol{A}_j$, $j=1,\ldots,J$ carry some simple
relationships. We present the result in the following lemma.

\begin{lemma}
Suppose the covariance matrix $\boldsymbol{K}_w$ satisfies
(\ref{equ:layered_correlation}). If $\boldsymbol{K}_w\succ
\boldsymbol{0}$ and $\boldsymbol{A}_j\succeq \boldsymbol{0}$,
$j=1,\ldots,J$, then
\begin{small}\begin{equation}
\left\{\begin{array}{l}\left(\boldsymbol{\underline{K}}_{\{1,\ldots,M_1^1\}}+\boldsymbol{A}_1\right)^{-1}=\sum_{i=1}^{m_1}\left(\boldsymbol{K}_i+\boldsymbol{A}_1\right)^{-1} \\
\left(\boldsymbol{\underline{K}}_{\{1,\ldots,M_1^{j}\}}+\boldsymbol{A}_j\right)^{-1}=\left(\boldsymbol{\underline{K}}_{\{1,\ldots,M_1^{j-1}\}}+\boldsymbol{A}_j\right)^{-1}
+\sum_{i=1}^{m_j}\left(\boldsymbol{K}_{M_1^{j-1}+i}+\boldsymbol{A}_j\right)^{-1},\quad j=2,\ldots,J\end{array}.\right.
\label{equ:A_K_relation}
\end{equation}\end{small}\hspace{-1mm}
\label{lemma:A_K_relation}
\end{lemma}
\begin{proof}
See Appendix \ref{appendix:proof_lemma_A_K_relation}.
\end{proof}

Equ. (\ref{equ:proportionality}) together with
(\ref{equ:A_K_relation}) fully characterizes the relationships
among $\boldsymbol{\underline{K}}_{\{1,\ldots,M_1^{j}\}}$ and
$\boldsymbol{A}_j$, $j=1,\ldots,J$. Note that given
$\{\boldsymbol{K}_l:l=1,\ldots,L\}$,
(\ref{equ:proportionality})-(\ref{equ:A_K_relation}) can be
interpreted as a set of functions $\boldsymbol{A}_j,j=2,\ldots,J$,
and $\boldsymbol{\underline{K}}_{\{1,\ldots,M_1^j\}},j=1,\ldots,J$
over $\boldsymbol{A}_1$. In other words, $\boldsymbol{A}_1$ is the
only free covariance matrix under
(\ref{equ:proportionality})-(\ref{equ:A_K_relation}). The problem
for obtaining $\boldsymbol{K}_w$ is now reduced to determine the
correlation matrix $\boldsymbol{A}_1$ and
$\{\boldsymbol{K}_l:l=1,\ldots,L\}$. In fact, the covariance
matrices $\{\boldsymbol{K}_l:l=1,\ldots,L\}$ can be determined by
the individual side distortion constraints. On the other hand, the
correlation matrix $\boldsymbol{A}_1$ can be determined by the
central distortion constraint.

We now revisit the side and central distortion constraints.
Suppose the distortion constraints
$(\boldsymbol{D}_1,\ldots,\boldsymbol{D}_L,\boldsymbol{D}_0)$ are
achieved with equality using the Gaussian description scheme. By
applying (\ref{equ:MMSE_x2}), we have
\begin{small}\begin{eqnarray}
&& \boldsymbol{K}_l=(\boldsymbol{D}_l^{-1}-\boldsymbol{K}_x^{-1})^{-1},\quad l=1,\ldots,L \nonumber\\
&&
\underline{\boldsymbol{K}}_w=(\boldsymbol{D}_0^{-1}-\boldsymbol{K}_x^{-1})^{-1}.
\label{equ:virtual_channel_distortion}\end{eqnarray}\end{small}\hspace{-1mm}
If there exists a positive definite solution $\boldsymbol{A}_1$ to
(\ref{equ:proportionality})-(\ref{equ:virtual_channel_distortion}),
then the distortion constraints are met with equality. The
remaining work is to check if the matrix  $\boldsymbol{K}_w$
constructed from the solutions of
(\ref{equ:proportionality})-(\ref{equ:virtual_channel_distortion})
is positive definite. It turns out that as long as
$\boldsymbol{A}_1$ is a solution to
(\ref{equ:proportionality})-(\ref{equ:A_K_relation}) where
$\{\boldsymbol{K}_l\succ\boldsymbol{0}:l=1,\ldots, L\}$ and
$\underline{\boldsymbol{K}}_w\succ\boldsymbol{0}$ are constant
matrices, the resulting $\boldsymbol{K}_w$ is always positive
definite. We state this formally below.

\begin{lemma}
Let $\underline{\boldsymbol{K}}_w\succ\boldsymbol{0}$ and
$\boldsymbol{K}_l\succ\boldsymbol{0}$, $l=1,\ldots,L$ be constant
matrices in (\ref{equ:proportionality})-(\ref{equ:A_K_relation}).
If for some $\boldsymbol{A}_1\succ\boldsymbol{0}$,
(\ref{equ:proportionality})-(\ref{equ:A_K_relation}) are true and
$\boldsymbol{0}\prec\alpha_1\boldsymbol{A}_1\prec\alpha_2\boldsymbol{A}_2\prec\ldots
\prec\alpha_J\boldsymbol{A}_J$, then the matrix $\boldsymbol{K}_w$
constructed using (\ref{equ:layered_correlation}) is positive
definite. \label{lemma:Kw_positive}
\end{lemma}

\begin{proof}
See Appendix \ref{lemma:Kw_positive_proof}.
\end{proof}

We summarize the result in the following proposition.
\begin{proposition}
Given distortion constraints
$(\boldsymbol{D}_1,\ldots,\boldsymbol{D}_L,\boldsymbol{D}_0)$. If
there exists a solution $\boldsymbol{A}_1^{\ast}$ to
(\ref{equ:proportionality})-(\ref{equ:virtual_channel_distortion}),
then the Gaussian description scheme with $\boldsymbol{K}_w$
constructed using (\ref{equ:layered_correlation}) achieves the
optimal weighted sum rate. The optimal $\boldsymbol{N}_j$ in
(\ref{equ:lower_bound}) is given as
$\boldsymbol{N}_j=(\boldsymbol{A}_j(\boldsymbol{A}_1^{\ast})^{-1}-\boldsymbol{K}_x^{-1})^{-1}$,
$j=1,\ldots,J$. \label{proposition:optimal_weighted_sumrate}
\end{proposition}

\subsection{Rate Region for the Scalar Gaussian Source}
In this subsection, we use Theorem 2.3 to prove that for the
scalar Gaussian case, the outer bound on the weighted sum-rate
provided by (\ref{equ:lower_bound}) is tight. This completely
characterizes the rate region for the scalar Gaussian case and
parallels a recent result of Chen \cite{Chen09MDCRateRegion}.

\begin{theorem}
For a scalar Gaussian source and individual and central distortion
constraints, the complete rate region is given by
(\ref{equ:lower_bound}) and (\ref{equ:lower_bound_sumRate}).
\label{theorem:rate_region_scalar}
\end{theorem}

\begin{proof}
See Section \ref{section:scalar_Gaussian}.
\end{proof}

\begin{remark} It should be noted from Theorem \ref{theorem:rate_region_scalar} and (\ref{equ:lower_bound})-(\ref{equ:lower_bound_sumRate}), that
the scalar Gaussian rate region is obtained by solving a single
maximization problem, whereas the approach of
\cite{Chen09MDCRateRegion} requires the solution of a min-max
optimization problem.
\end{remark}

\section{Proof of Theorem \ref{theorem:outer_bound}}
\label{section:outer_bound}

Before presenting the proof of Theorem 2.2, we need the following
lemma:

\begin{lemma}
Assume that $\boldsymbol{v}$ and $\boldsymbol{x}^{n}$ are
arbitrarily distributed random variables (which might be
correlated) and assume that given $\boldsymbol{v}$,
$\boldsymbol{x}^{n}$ has a density. Denote the dimensionality of
$\boldsymbol{x}^{n}$ as $nN$. Let
$\{\boldsymbol{z}_1[m]\}_{m=1}^n$ and
$\{\boldsymbol{z}_2[m]\}_{m=1}^n$ be two i.i.d. random Gaussian
vector processes with marginal distributions
$\mathcal{N}(\boldsymbol{0},\boldsymbol{N}_1)$ and
$\mathcal{N}(\boldsymbol{0},\boldsymbol{N}_2)$, respectively. Both
$\boldsymbol{N}_1$ and $\boldsymbol{N}_2$ are of size $N\times N$.
Let
$\boldsymbol{z}_1^{n}=[\boldsymbol{z}_1[1]^t,\ldots,\boldsymbol{z}_1[n]^t]^t$.
In a similar manner, we can define $\boldsymbol{z}_2^{n}$. Both
$\boldsymbol{z}_1^{n}$ and $\boldsymbol{z}_2^{n}$ are independent
of $\boldsymbol{v}$ and $\boldsymbol{x}^{n}$. Let $\mu_1
> \mu_2 > 0$ and $\boldsymbol{0}\prec\mu_1\boldsymbol{N}_1 \prec
\mu_2\boldsymbol{N}_2$. Then there is
\begin{small}\begin{eqnarray}
&&\hspace{-6mm}\mu_2 h(\boldsymbol{x}^{n}+\boldsymbol{z}_2^{n}|\boldsymbol{v})-\mu_1 h(\boldsymbol{x}^{n}+\boldsymbol{z}_1^{n}|\boldsymbol{v})
+(\mu_1-\mu_2)h(\boldsymbol{x}^{n}|\boldsymbol{v}) \nonumber\\
&& \leq \frac{n\mu_1}{2} \log\frac{(\mu_1-\mu_2)^N|\boldsymbol{N}_2|}{\mu_1^N|\boldsymbol{N}_2-\boldsymbol{N}_1|}
-\frac{n\mu_2}{2}\log\frac{(\mu_1-\mu_2)^N|\boldsymbol{N}_1|}{\mu_2^N|\boldsymbol{N}_2-\boldsymbol{N}_1|},
\label{equ:extreme_inequality}
\end{eqnarray}\end{small}\hspace{-1mm}
where the equality holds if $\boldsymbol{x}^{n}$ and
$\boldsymbol{v}$ are jointly Gaussian and satisfy
\begin{small}\begin{equation}
\textrm{Cov}[\boldsymbol{x}^n|\boldsymbol{v}]=\boldsymbol{I}_n\otimes (\mu_1-\mu_2)\boldsymbol{N}_2(\mu_2\boldsymbol{N}_2-\mu_1\boldsymbol{N}_1)^{-1}\boldsymbol{N}_1,
\label{equ:covariance_condition}
\end{equation}\end{small}\hspace{-1mm}
where $\otimes$ denotes the Kronecker product \cite{Horn90Matrix}.
\label{lemma:EPI}
\end{lemma}

\begin{proof}See Appendix \ref{appendix:Proof of Lemma:EPI}.  \end{proof}

When the problem specializes to the scalar source case (i.e.
$N=1$), an alternative upper bound of the right hand side of
(\ref{equ:extreme_inequality}) was derived by Chen
\cite{Chen09MDCRateRegion}. The main difference is that the upper
bound in \cite{Chen09MDCRateRegion} requires solving an
optimization problem, and does not admit a simple closed-form
expression.

We now proceed to present the argument for the outer bound. We
first consider the case $J>1$. We define $J$ i.i.d. random
Gaussian processes
$\boldsymbol{z}_i^{n}=\{\boldsymbol{z}_i[m]\}_{m=1}^{n}$ with
marginal distributions
$\mathcal{N}(\boldsymbol{0},\boldsymbol{N}_j)$, $j=1,\ldots,J$,
respectively. Let $C_l=f^{(n)}_l(\boldsymbol{x}^n)$,
$l=1,\ldots,L$, be the discrete random variables. Suppose that
$\boldsymbol{z}_1^{n}$, $\ldots$, $\boldsymbol{z}_J^{n}$ are
independent of $\boldsymbol{x}^{n}$ and $C_l$, $l=1,\ldots,L$.
Next we introduce $J$ processes
$\boldsymbol{y}_j^n=(\boldsymbol{y}_j[1]^t,\ldots,\boldsymbol{y}_j[n]^t)^t$,
$j=1,\ldots,J$ by
\begin{small}\begin{equation}
\boldsymbol{y}_j[m]=\boldsymbol{x}[m]+\boldsymbol{z}_j[m], \quad m=1,\ldots,n \textrm{ and } j=1,\ldots,J.
\end{equation}\end{small}\hspace{-1mm}
It follows that each sequence $\{\boldsymbol{y}_j[m]\}$ is an
i.i.d. process with marginal distribution
$\mathcal{N}(\boldsymbol{0},\boldsymbol{K}_{y_j})$, where
$\boldsymbol{K}_{y_j}=\boldsymbol{K}_x+\boldsymbol{N}_j$,
$j=1,\ldots,J$. The following sequence of inequalities provides a
lower bound to the weighted sum rate
(\ref{equ:weighted_sume_optimization2}):
\begin{small}\begin{equation}
\hspace{-100mm}n\alpha_0R_1+\sum_{j=1}^{J}n\alpha_j
\left(\sum_{i=1}^{m_j}R_{M_1^{j-1}+i}\right)
\nonumber\vspace{-3mm}
\end{equation}
\begin{equation}
\hspace{-88mm}\geq \alpha_0H(C_1)+\sum_{j=1}^{J}\alpha_j \left(\sum_{i=1}^{m_j}H\left(C_{M_1^{j-1}+i}\right)\right) \nonumber
\vspace{-3mm}
\end{equation}
\begin{equation}
\hspace{-26mm}\stackrel{(a)}{\geq}
\alpha_0\left[H(C_1)-H(C_1|\boldsymbol{x}^n)\right]+\sum_{j=1}^{J}\alpha_j
\Bigg(\sum_{i=1}^{m_j}H\left(C_{M_1^{j-1}+i}\right)-\Big[H(C_1,\ldots,C_{M_1^{j-1}}|\boldsymbol{y}_j^n)+
\nonumber\vspace{-3mm}
\end{equation}
\begin{equation}
\hspace{-72mm}\sum_{i=1}^{m_j}H\left(C_{M_1^{j-1}+i}|\boldsymbol{y}_j^n\right)-H\left(C_1,\ldots,C_{M_1^j}|\boldsymbol{y}_j^n\right) \Big]\Bigg)
\nonumber\vspace{-3mm}
\end{equation}
\begin{equation}
 \hspace{4mm}= \alpha_0
\left[h(\boldsymbol{x}^n)-h(\boldsymbol{x}^n|C_1)\right]+\sum_{j=1}^{J}\alpha_j
\sum_{i=1}^{m_j}I(\boldsymbol{y}_j^n;C_{M_1^{j-1}+i})
+\alpha_J\left[H\left(C_1,\ldots,C_{L}|\boldsymbol{y}_J^n\right)-H\left(C_1,\ldots,C_L|\boldsymbol{x}^n\right)\right]+\nonumber\vspace{-3mm}
\end{equation}
\begin{equation}
 \hspace{9mm}\sum_{j=2}^{J}\left[\alpha_{j-1}H\left(C_1,\ldots,C_{M_1^{j-1}}|\boldsymbol{y}_{j-1}^n\right)-\alpha_jH\left(C_1,\ldots,C_{M_1^{j-1}}|\boldsymbol{y}_{j}^n\right)
-(\alpha_{j-1}-\alpha_j)H\left(C_1,\ldots,C_{M_1^{j-1}}|\boldsymbol{x}^n\right)\right] \nonumber\vspace{-3mm}
\end{equation}
\begin{equation}
\hspace{-48mm}=\alpha_0
\left[h(\boldsymbol{x}^n)-h(\boldsymbol{x}^n|C_1)\right]+
\sum_{j=1}^{J}\alpha_j
\sum_{i=1}^{m_j}\left(h(\boldsymbol{y}_j^n)-h\left(\boldsymbol{y}_j^n|C_{M_1^{j-1}+i}\right)\right)+
\nonumber\vspace{-3mm}
\end{equation}
\begin{equation}
\hspace{5mm}\sum_{j=2}^{J}\left[\alpha_{j-1}h\left(\boldsymbol{y}_{j-1}^n|C_1,\ldots,C_{M_1^{j-1}}\right)-\alpha_jh\left(\boldsymbol{y}_{j}^n|C_1,\ldots,C_{M_1^{j-1}}\right)
-(\alpha_{j-1}-\alpha_j)h\left(\boldsymbol{x}^n|C_1,\ldots,C_{M_1^{j-1}}\right)\right] \nonumber\vspace{-3mm}
\end{equation}
\begin{equation}
\hspace{-43mm}+\alpha_J\left[h\left(\boldsymbol{y}_J^n|C_1,\ldots,C_{L}\right)-h\left(\boldsymbol{x}^n|C_1,\ldots,C_L\right)\right]
+ \alpha_1h(\boldsymbol{x}^n)-\alpha_1h(\boldsymbol{y}_1^n).
\label{equ:lower_bound_2}
\end{equation}\end{small}\hspace{-1mm}
The construction of the inequality $(a)$ is crucial in the
outer-bound derivation. For each distinct weighting factor, we
essentially introduce an auxiliary Gaussian process.

We now consider deriving a lower bound to
(\ref{equ:lower_bound_2}). As $\boldsymbol{x}^n$ and
$\boldsymbol{y}_j^n$, $j=1,\ldots,L$ are Gaussian vectors, we
easily obtain
\begin{small}\begin{eqnarray}
&&h(\boldsymbol{x}^n)=\frac{1}{2}\log(2\pi e)^{Nn}|\boldsymbol{K}_x|^n, \nonumber\\
&&h(\boldsymbol{y}_j^n)=\frac{1}{2}\log(2\pi e)^{Nn}|\boldsymbol{K}_x+\boldsymbol{N}_j|^n, \quad j=1,\ldots,J.
\label{equ:entropy_inequlity_1}
\end{eqnarray}\end{small}\hspace{-1mm}
We will also use some entropy-related inequalities developed in
\cite{Wang07VectorMDC}, which are
\begin{small}\begin{equation}
\hspace{-78mm} h(\boldsymbol{x}^n|C_1)\leq \frac{1}{2}\log(2\pi
e)^{Nn}|\boldsymbol{D}_1|^n, \label{equ:entropy_inequlity_2}
\vspace{-3mm}
\end{equation}
\begin{equation}
h(\boldsymbol{y}_j^n|C_{M_1^{j-1}+i})\leq \frac{1}{2}\log(2\pi
e)^{Nn}|\boldsymbol{D}_{M_1^{j-1}+i}+\boldsymbol{N}_j|^n, \quad
i=1,\ldots,m_j \textrm{ and } j=1,\ldots,J,
\label{equ:entropy_inequlity_3}
\vspace{-3mm}
\end{equation}
\begin{equation}
\hspace{-37mm}h\left(\boldsymbol{y}_J^n|C_1,\ldots,C_{L}\right)-h\left(\boldsymbol{x}^n|C_1,\ldots,C_L\right)\geq
\frac{n}{2}\log\frac{|\boldsymbol{D}_0+\boldsymbol{N}_J|}{|\boldsymbol{D}_0|}.
\label{equ:entropy_inequlity_4}
\end{equation}\end{small}\hspace{-1mm}
By using Lemma \ref{lemma:EPI}, the remaining quantities in
(\ref{equ:lower_bound_2}) can be lower-bounded as
\begin{small}\begin{eqnarray}
&& \alpha_{j-1}h\left(\boldsymbol{y}_{j-1}^n|C_1,\ldots,C_{M_1^{j-1}}\right)-\alpha_jh\left(\boldsymbol{y}_{j}^n|C_1,\ldots,C_{M_1^{j-1}}\right)
-(\alpha_{j-1}-\alpha_j)h\left(\boldsymbol{x}^n|C_1,\ldots,C_{M_1^{j-1}}\right) \nonumber\\
&&\geq -\frac{n\alpha_{j-1}}{2}\log\frac{(\alpha_{j-1}-\alpha_j)^N|\boldsymbol{N}_j|}{\alpha_{j-1}^N|\boldsymbol{N}_j-\boldsymbol{N}_{j-1}|}+
\frac{n\alpha_j}{2}\log\frac{(\alpha_{j-1}-\alpha_j)^N|\boldsymbol{N}_{j-1}|}{\alpha_j^N|\boldsymbol{N}_j-\boldsymbol{N}_{j-1}|}, \quad j=2,\ldots,J, \label{equ:entropy_inequlity_5}
\end{eqnarray}\end{small}\hspace{-1mm}
where
$\boldsymbol{0}\prec\alpha_1\boldsymbol{N}_1\prec\ldots\prec\alpha_J\boldsymbol{N}_J.$
The equalities hold in (\ref{equ:entropy_inequlity_5}) if
$C_1,\ldots, C_L$ and $\boldsymbol{x}^n$ are jointly Gaussian with
conditional covariance matrices
\begin{small}\begin{equation}
\textrm{Cov}[\boldsymbol{x}^n|C_1,\ldots,C_{M_1^{j-1}}]
=\boldsymbol{I}_n\otimes (\alpha_{j-1}-\alpha_j)\boldsymbol{N}_j(\alpha_j\boldsymbol{N}_j-\alpha_{j-1}\boldsymbol{N}_{j-1})^{-1}\boldsymbol{N}_{j-1},\quad j=2,\ldots,J.
\label{equ:entropy_inequlity_5_equCon}
\end{equation}\end{small}\hspace{-1mm}
Plugging
(\ref{equ:entropy_inequlity_1})-(\ref{equ:entropy_inequlity_5})
into (\ref{equ:lower_bound_2}) produces the expression
(\ref{equ:lower_bound}).

For the case $J=1$ (i.e., the sum-rate case), we can follow the
same derivation steps as those for $J>1$. The argument for this
special case is actually the same as that in
\cite{Wang07VectorMDC}. An alternative derivation is to start with
the expression of the outer bound for $J>1$, and let
$\alpha_2,\ldots,\alpha_J$ approach $\alpha_1$ in the expression.
The proof is complete.

\section{Proof of Theorem \ref{theorem:optimal_weight_sumRate}}
\label{section:Opt_weight_sumRate}

In this section we study the optimality of the Gaussian
description scheme in achieving the outer bound
(\ref{equ:lower_bound}). We identify the optimality conditions by
investigating the inequalities used in the outer-bound derivation
under the Gaussian description scheme.

Similarly to the derivation of the outer bound, we introduce $J$
auxiliary Gaussian random vectors $\boldsymbol{z}_j\sim
\mathcal{N}(\boldsymbol{0},\boldsymbol{N}_j)$, $j=1,\ldots,J$,
independent of $\boldsymbol{x}$ and all $\boldsymbol{w}_l$'s. As
for (\ref{equ:entropy_inequlity_5}), we put a partial ordering
constraint on $\boldsymbol{N}_j$, $l=1,\ldots,J$:
\begin{small}\begin{equation}
\boldsymbol{0}\prec\alpha_1\boldsymbol{N}_1\prec\ldots\prec\alpha_J\boldsymbol{N}_J. \label{equ:N_inequality}
\end{equation}\end{small}\hspace{-1mm}
To simplify the discussion, we introduce $\alpha_{J+1}=0$. Letting
$\boldsymbol{y}_j=\boldsymbol{x}+\boldsymbol{z}_j$,
$j=1,\ldots,J$, we obtain the following lower bound to
(\ref{equ:weighted_sume_optimization2}) under the Gaussian
description scheme:
\begin{small}
\begin{equation}
\hspace{-100mm}\alpha_0R_1+\sum_{j=1}^J\alpha_j\left(\sum_{i=1}^{m_j}R_{M_1^{j-1}+i}\right)\nonumber
\vspace{-3mm}\end{equation}
\begin{equation}
\hspace{-92mm}=\alpha_0R_1+\sum_{j=1}^J(\alpha_j-\alpha_{j+1})\left( \sum_{i=1}^{M_1^j}R_{i}\right)
\nonumber
\vspace{-3mm}\end{equation}
\begin{equation}
\hspace{-29mm}\stackrel{(a)}{=}\alpha_0\left(h(\boldsymbol{u}_1)-h(\boldsymbol{u}_1|\boldsymbol{x})\right)+
\sum_{j=1}^J(\alpha_j-\alpha_{j+1})\left( \sum_{i=1}^{M_1^{j}}h\left(\boldsymbol{u}_{i}\right)
-h\left(\boldsymbol{u}_{1},\ldots,\boldsymbol{u}_{M_1^{j}}|\boldsymbol{x}\right)\right)\nonumber
\vspace{-3mm}\end{equation}
\begin{equation}
\hspace{-25mm}= \alpha_0I\left(\boldsymbol{u}_1;\boldsymbol{x}\right)
-\sum_{j=1}^{J}(\alpha_j-\alpha_{j+1})h\left(\boldsymbol{u}_{1},\ldots,\boldsymbol{u}_{M_1^{j}}|\boldsymbol{x}\right)
+\sum_{j=1}^J\alpha_j\left( \sum_{i=1}^{m_j}h\left(\boldsymbol{u}_{M_1^{j-1}+i}\right)\right) \nonumber
\vspace{-3mm}\end{equation}
\begin{equation}
\hspace{-27mm}\stackrel{(b)}{\geq}\alpha_0I\left(\boldsymbol{u}_1;\boldsymbol{x}\right)
-\sum_{j=1}^{J}(\alpha_j-\alpha_{j+1})h\left(\boldsymbol{u}_{1},\ldots,\boldsymbol{u}_{M_1^{j}}|\boldsymbol{x}\right)
+\sum_{j=1}^J\alpha_j\left( \sum_{i=1}^{m_j}h\left(\boldsymbol{u}_{M_1^{j-1}+i}\right) \right.\nonumber
\vspace{-3mm}\end{equation}
\begin{equation} \hspace{-25mm} \left.-\Bigg[
h\left(\boldsymbol{u}_1,\ldots,\boldsymbol{u}_{M_1^{j-1}}|\boldsymbol{y}_j\right)
+\sum_{i=1}^{m_j}h\left(\boldsymbol{u}_{M_1^{j-1}+i}|\boldsymbol{y}_j\right)
-h\left(\boldsymbol{u}_1,\ldots,\boldsymbol{u}_{M_1^{j}}|\boldsymbol{y}_j \right) \Bigg]\right) \nonumber
\vspace{-3mm}\end{equation}
\begin{equation}
\hspace{-10mm}=\alpha_0I\left(\boldsymbol{u}_1;\boldsymbol{x}\right)+
\sum_{j=1}^J\alpha_j\left(
\sum_{i=1}^{m_j}I\left(\boldsymbol{u}_{M_1^{j-1}+i};\boldsymbol{y}_j\right)\right)
+\alpha_J\left[h(\boldsymbol{y}_J|\boldsymbol{u}_1,\ldots,\boldsymbol{u}_L)-h(\boldsymbol{x}|\boldsymbol{u}_1,\ldots,\boldsymbol{u}_L)\right] \nonumber
\vspace{-3mm}\end{equation}
\begin{equation}
\hspace{9mm}+\sum_{j=2}^{J}\left[\alpha_{j-1}h\left(\boldsymbol{y}_{j-1}|\boldsymbol{u}_1,\ldots,\boldsymbol{u}_{M_1^{j-1}}\right)
-\alpha_jh\left(\boldsymbol{y}_{j}|\boldsymbol{u}_1,\ldots,\boldsymbol{u}_{M_1^{j-1}}\right)
-(\alpha_{j-1}-\alpha_j)h\left(\boldsymbol{x}|\boldsymbol{u}_1,\ldots,\boldsymbol{u}_{M_1^{j-1}}\right)\right] \nonumber
\vspace{-3mm}\end{equation}
\begin{equation}
\hspace{-110mm}+\alpha_1h(\boldsymbol{x})-\alpha_1h(\boldsymbol{y}_1),
\label{equ:optimality_condition_1}
\vspace{-3mm}\end{equation}\end{small}\hspace{-1mm}
where step $(a)$ follows from Lemma \ref{lemma:rate_region_inner}.

We now study the inequality (b) in the derivation of
(\ref{equ:optimality_condition_1}). It is straightforward that if
\begin{small}\begin{equation}
h\left(\boldsymbol{u}_1,\ldots,\boldsymbol{u}_{M_1^{j-1}}|\boldsymbol{y}_j\right)
+\sum_{i=1}^{m_j}h\left(\boldsymbol{u}_{M_1^{j-1}+i}|\boldsymbol{y}_j\right)
-h\left(\boldsymbol{u}_1,\ldots,\boldsymbol{u}_{M_1^{j}}|\boldsymbol{y}_j \right)=0, \quad j=1,\ldots,J,
\label{equ:cond_independence}
\end{equation}\end{small}\hspace{-1mm}
then (b) is actually an equality. The fact that
(\ref{equ:cond_independence}) hold implies that conditioned on
$\boldsymbol{y}_j$,
$\boldsymbol{u}_{M_1^{j-1}+1},\cdots,\boldsymbol{u}_{M_1^j}$, and
$[\boldsymbol{u}_1^t,\ldots,\boldsymbol{u}_{M_1^{j-1}}^t]^t$ are
independent. In other words, the $m_j$ descriptions associated
with $\alpha_j$, and the set of descriptions associated with
$\{\alpha_i,i<j\}$, are conditionally independent given
$\boldsymbol{y}_j$. Further, the conditional independencies
(\ref{equ:cond_independence}) exhibit a layered structure. The
conditional independency w.r.t. $\boldsymbol{y}_j$ only refers to
those descriptions with weighting factors no larger than
$\alpha_j$. The number of layered conditional independency is
determined by the number of distinct weighting factors. Note that
all the random variables involved in
(\ref{equ:optimality_condition_1}) are Gaussian vectors. The
condition (\ref{equ:cond_independence}) can be equivalently
described by constraints on their covariance matrices. We present
these constraints in the following proposition.

\begin{proposition}
There exist a set of matrices
$(\boldsymbol{N}_1,\ldots,\boldsymbol{N}_{J})$ such that
(\ref{equ:N_inequality}) holds and (\ref{equ:cond_independence})
is true if the covariance matrix $\boldsymbol{K}_w$ satisfies the
following three conditions
\begin{itemize}
    \item equation (\ref{equ:layered_correlation}) is true;
    \item the covariance matrix $\boldsymbol{N}_j$ takes the
        form
\begin{small}\begin{equation}
\boldsymbol{N}_j=(\boldsymbol{A}_j^{-1}-\boldsymbol{K}_x^{-1})^{-1}, \quad j=1,\ldots,J.
\label{equ:A_N_relation}
\end{equation}\end{small}\hspace{-1mm}
    \item The correlation matrices $\boldsymbol{A}_j$,
        $j=1,\ldots,J$ satisfy a partial ordering constraint:
        \begin{small}\begin{eqnarray}\boldsymbol{0}\prec
        \alpha_1\boldsymbol{A}_1
        \prec\alpha_2\boldsymbol{A}_2\prec\ldots\prec
        \alpha_J\boldsymbol{A}_J\prec
        \alpha_J\boldsymbol{K}_x.
        \label{equ:A_ordering}
          \end{eqnarray}\end{small}\hspace{-1mm}
\end{itemize}
\label{proposition:A_N_relation}
\end{proposition}

\begin{proof}

We first derive the conditions such that
(\ref{equ:cond_independence}) is true. Then we consider the
partial ordering constraint (\ref{equ:N_inequality}) to obtain the
additional conditions.

Note that the $J$ conditional independencies in
(\ref{equ:cond_independence}) have the same structure. We can
focus on one particular case. The argument for other cases are the
same. For a particular $j$, we use similar derivation steps as in
the proof of \cite[Proposition 2]{Wang07VectorMDC} to obtain the
conditions. We find that (\ref{equ:cond_independence}) holds for
each $j$ when
\begin{small}\begin{eqnarray}
&&\mathbb{E}\left[\boldsymbol{w}_{M_1^{j-1}+i}\boldsymbol{w}_{k}\right]=-\boldsymbol{A}_j,
\quad \forall 1\leq k < M_1^{j-1}+i,\textrm{ }
i=1,\ldots,m_j \nonumber\\
&& \boldsymbol{0}\prec\boldsymbol{A}_j\prec\boldsymbol{K}_x \nonumber \\
&&\boldsymbol{N}_j=(\boldsymbol{A}_j^{-1}-\boldsymbol{K}_x^{-1})^{-1}\nonumber.
\end{eqnarray}\end{small}\hspace{-1mm}
Proposition 2 in \cite{Wang07VectorMDC} actually considered the
conditional independency for $J=1$ (the sum-rate case).


Next we show that (\ref{equ:A_N_relation}) and
(\ref{equ:A_ordering}) together are sufficient to produce
(\ref{equ:N_inequality}). By using Lemma
\ref{lemma:ordering_relation}, we have
\begin{small}\begin{equation}
\hspace{-40mm}\alpha_{j}\boldsymbol{A}_j\prec\alpha_{j+1}\boldsymbol{A}_{j+1}\Leftrightarrow
\alpha_{j+1}\boldsymbol{A}_j^{-1}\succ\alpha_{j}\boldsymbol{A}_{j+1}^{-1}
\nonumber\vspace{-3mm}
\end{equation}
\begin{equation}
\hspace{19mm}\Rightarrow  \alpha_{j+1}\boldsymbol{A}_j^{-1}+(\alpha_j-\alpha_{j+1})\boldsymbol{K}_x^{-1}\succ \alpha_{j}\boldsymbol{A}_{j+1}^{-1} \nonumber
\vspace{-3mm}\end{equation}
\begin{equation}
\hspace{19mm}\Leftrightarrow  \alpha_{j+1}(\boldsymbol{A}_j^{-1}-\boldsymbol{K}_x^{-1})\succ \alpha_{j}(\boldsymbol{A}_{j+1}^{-1}-\boldsymbol{K}_x^{-1}) \nonumber
\vspace{-3mm}\end{equation}
\begin{equation}
\hspace{19mm}\Leftrightarrow  \alpha_{j}\boldsymbol{N}_j\prec
\alpha_{j+1}\boldsymbol{N}_{j+1},\quad j=1,\ldots,J-1. \nonumber
\end{equation}\end{small}\hspace{-1mm}
The proof is complete.
\end{proof}

We consider the matrix $\boldsymbol{K}_w$ satisfying the
conditions in Proposition \ref{proposition:A_N_relation}. We
proceed to recognize the additional conditions such that
(\ref{equ:optimality_condition_1}) reaches the outer bound
(\ref{equ:lower_bound}). Again since all the random variables
involved in (\ref{equ:optimality_condition_1}) are Gaussian
vectors, knowing their covariance matrices is sufficient to
characterize the expression. We list the distortion conditions as
\begin{small}\begin{eqnarray}
&& \textrm{Cov}[\boldsymbol{x}|\boldsymbol{u}_l]=\boldsymbol{D}_l, \quad l=1,\ldots,L, \label{equ:optimality_condition_sideDis} \\
&& \textrm{Cov}\left[\boldsymbol{x}|\boldsymbol{u}_1,\ldots,\boldsymbol{u}_{M_1^{j-1}} \right]=
(\boldsymbol{K}_x^{-1}+ \frac{\alpha_j}{\alpha_{j-1}-\alpha_j}\boldsymbol{A}_{j-1}^{-1}-\frac{\alpha_{j-1}}{\alpha_{j-1}-\alpha_j}\boldsymbol{A}_{j}^{-1})^{-1},\quad  j=2,\ldots,J,
\label{equ:optimality_condition_midDis}\\
&& \textrm{Cov}[\boldsymbol{x}|\boldsymbol{u}_1,\ldots,\boldsymbol{u}_L]=\boldsymbol{D}_0.
\label{equ:optimality_condition_centralDis}
\end{eqnarray}\end{small}\hspace{-1mm}
The above distortion conditions are derived from
(\ref{equ:entropy_inequlity_1})-(\ref{equ:entropy_inequlity_5_equCon})
and (\ref{equ:A_N_relation}). In particular,
(\ref{equ:optimality_condition_midDis}) is obtained by combining
(\ref{equ:entropy_inequlity_5_equCon}) and
(\ref{equ:A_N_relation}). (\ref{equ:optimality_condition_sideDis})
and (\ref{equ:optimality_condition_centralDis}) are for the
individual side distortion constraints and the central distortion
constraint, respectively. They correspond to
(\ref{equ:entropy_inequlity_2})-(\ref{equ:entropy_inequlity_4}).
By using (\ref{equ:A_N_relation}) and
(\ref{equ:optimality_condition_sideDis})-(\ref{equ:optimality_condition_centralDis}),
it can be shown that (\ref{equ:optimality_condition_1}) gives the
same expression as the outer bound (\ref{equ:lower_bound}).

It is now clear that (\ref{equ:layered_correlation}),
(\ref{equ:A_ordering}) and
(\ref{equ:optimality_condition_sideDis})-(\ref{equ:optimality_condition_centralDis})
are optimality conditions on the covariance matrix
$\boldsymbol{K}_w$ for the weighted sum rate. When a proper
$\boldsymbol{K}_w$ can be constructed satisfying the optimality
conditions, the outer bound (\ref{equ:lower_bound}) is tight under
the expression (\ref{equ:A_N_relation}). From (\ref{equ:MMSE_x2}),
the condition (\ref{equ:optimality_condition_midDis}) can be
rewritten as
\begin{small}\begin{equation}
\boldsymbol{\underline{K}}_{\{1,\ldots,M_1^j\}}^{-1}=\frac{\alpha_{j+1}}{\alpha_j-\alpha_{j+1}}\boldsymbol{A}_j^{-1}-\frac{\alpha_j}{\alpha_j-\alpha_{j+1}}\boldsymbol{A}_{j+1}^{-1}, \quad j=1,\ldots,J-1.
\label{equ:proportionality2}
\end{equation}\end{small}\hspace{-1mm}
By using Lemma \ref{lemma:matrix_inversion} on
(\ref{equ:proportionality2}), we then arrive at the
\emph{proportionality} condition (\ref{equ:proportionality}). For
each $j=1,\ldots,J-1$, the expressions (\ref{equ:proportionality})
and (\ref{equ:proportionality2}) are equivalent when
$\boldsymbol{A}_j$ and $\boldsymbol{A}_{j+1}$ are positive
definite. However, the condition (\ref{equ:proportionality}) is
more powerful in that it can handle the case that
$\boldsymbol{A}_j\succeq \boldsymbol{0}$, $j=1,\ldots,J$. The
generality of (\ref{equ:proportionality}) over
(\ref{equ:proportionality2}) might be useful to extend Theorem
\ref{theorem:optimal_weight_sumRate} to the case that the central
distortion is loose. In our work, we will not discuss this case.
The proof is complete.

\section{Rate region for Scalar Gaussian Source}
\label{section:scalar_Gaussian} In this section we consider the
rate region of the multiple description problem with individual and central
receivers for a scalar Gaussian source
$x\sim\mathcal{N}(0,\sigma_x^2)$. Denote the individual and
central distortion constraints as $d_l$, $l=1,\ldots,L$, and
$d_0$, where $0<d_0<d_l<\sigma_x^2$ for all $l=1,\ldots,L$. We
construct the Gaussian test channel with $w_l$, $l=1,\ldots,L$,
such that
\begin{small}\begin{eqnarray}
    && k_l=\textrm{Cov}\left[w_l\right], \nonumber\\
    &&\mathbb{E}\left[w_{M_1^{j-1}+i}w_k \right]=-\sigma_j^2\quad \forall 1\leq k< M_1^{j-1}+i, \quad i=1,\ldots,m_j, \textrm{ and } j=1,\ldots,J.
\end{eqnarray}\end{small}\hspace{-1mm}
The correlation coefficient $\sigma_j^2$ corresponds to
$\boldsymbol{A}_j$ in (\ref{equ:layered_correlation}) defined for
general vector case.  With this Gaussian description scheme, we
find that the outer bound in \emph{Theorem}
\ref{theorem:outer_bound} is tight.

It is evident from Proposition
\ref{proposition:optimal_weighted_sumrate} that the conditions
(\ref{equ:proportionality}) and (\ref{equ:A_K_relation}) play an
important role in establishing optimality of a Gaussian test
channel. Note that for the scalar case,
(\ref{equ:proportionality}) is equivalent to
(\ref{equ:proportionality2}) when the correlation coefficients
$\sigma_j^2>0$, $j=1,\ldots,J$. We study the properties of
$\sigma_j^2$, $j=2,\ldots,J$ and
$\underline{k}_{\{1,\ldots,M_1^j\}}$, $j=1,\ldots,J$ as functions
of $\sigma_1^2$ through (\ref{equ:proportionality2}) and
(\ref{equ:A_K_relation}).

\begin{lemma}
Let $k_l>0$, $l=1,\ldots,L$, be constants. Define $\sigma_j^{2}$,
$j=2,\ldots,J$, and $\underline{k}_{\{1,\ldots,M_1^j\}}$,
$j=1,\ldots,J$ as functions of $\sigma_1^2$, expressed as
\begin{small}\begin{equation}
\left\{
\begin{array}{l}
\left(\underline{k}_{\{1,\ldots,M_1^1\}}+\sigma_1^2\right)^{-1}=\sum_{i=1}^{m_1}\left(\sigma_1^2+k_i\right)^{-1}\\
\left(\underline{k}_{\{1,\ldots,M_1^j\}}+\sigma_j^2\right)^{-1}=
\left(\underline{k}_{\{1,\ldots,M_1^{j-1}\}}+\sigma_j^2\right)^{-1}+\sum_{i=1}^{m_j}\left(\sigma_j^2+k_{M_1^{j-1}+i}\right)^{-1}, \quad j=2,\ldots, J
\end{array},
\right.\label{equ:A_K_relation_scalar}
\end{equation}\end{small}\hspace{-1mm}
\begin{small}\begin{equation}
\hspace{-50mm}\textrm{and}\quad\underline{k}_{\{1,\ldots,M_1^j\}}^{-1}=\frac{\alpha_{j+1}}{(\alpha_j-\alpha_{j+1})\sigma_j^2}-\frac{\alpha_j}{(\alpha_j-\alpha_{j+1})\sigma_{j+1}^2}, \quad j=1,\ldots,J-1.\label{equ:A_K_relation_scalar2}
\end{equation}\end{small}\hspace{-1mm}
Then there exists $\ddot{\sigma}_1^2>0$ such that the variables
$\sigma_l^2$, $l=2,\ldots,L$, are monotonically increasing over
$\sigma_1^2\in (0,\ddot{\sigma}_1^2)$, and the variables
$\underline{k}_{\{1,\ldots,M_1^j\}}$, $j=1,\ldots,J$, are
monotonically deceasing over $\sigma_1^2\in
(0,\ddot{\sigma}_1^2)$. For any $\sigma_1^2\in
(0,\ddot{\sigma}_1^2)$, there is
\begin{small}\begin{equation}
0<\alpha_1\sigma_1^2<\alpha_2\sigma_2^2<,\ldots,<\alpha_J\sigma_J^2.\label{equ:A_inequality_scalar}
\end{equation}\end{small}\hspace{-1mm}
Further,
\begin{small}\begin{equation}
\sigma_j^2\rightarrow \sigma_{j-1}^2 \textrm{ as
}\sigma_1^2\rightarrow 0, \quad \textrm{ where }  j=2,\ldots, J
\vspace{-3mm}\end{equation}
\begin{equation}
\hspace{-18mm}\underline{k}_w\rightarrow \left(\sum_{i=1}^L k_i^{-1}\right)^{-1}\textrm{ as }\sigma_1^2\rightarrow 0, \label{equ:k_w_bar_upBound}
\vspace{-3mm}\end{equation}
\begin{equation}
\hspace{-34mm} \underline{k}_w\rightarrow 0\textrm{ as }\sigma_1^2\rightarrow\ddot{\sigma}_1^2.
\end{equation}\end{small}\hspace{-1mm}

\label{lemma:scalar_monotonicity}
\end{lemma}
\begin{proof}
See Appendix \ref{appendix:proof_of_scalar_monotonicity} for the
proof.
\end{proof}

Upon establishing the result in \emph{Lemma}
\ref{lemma:scalar_monotonicity}, we are now ready to argue that
for the scalar Gaussian source, the multiple description scheme using Gaussian
descriptions achieves the outer bound in \emph{Theorem}
\ref{theorem:outer_bound}. In particular, we will show that when
either the central distortion constraint or the side distortion
constraint is loose, the outer bound
(\ref{equ:lower_bound})-(\ref{equ:lower_bound_sumRate}) is still
tight. Let
\begin{small}\begin{equation}
k_{l}=(d_l^{-1}-\sigma_x^{-2})^{-1},\quad l=1,\ldots,L \label{equ:k_l_scalar}
\end{equation}\end{small}\hspace{-1mm}
in \emph{Lemma} \ref{lemma:scalar_monotonicity}.  To simplify the
derivation, we use $\underline{k}_w(\sigma_1^2)$ and
$\sigma_j^2(\sigma_1^2)$, $j=2,\ldots,J$ to denote the functions
of $\underline{k}_w$ and $\sigma_j^2$, $j=2,\ldots,J$ over
$\sigma_1^2$. From (\ref{equ:k_w_bar_upBound}), we denote the
upper bound of $\underline{k}_w(\sigma_1^2)$ as
$\underline{k}^{up}_w=\left(\sum_{i=1}^Lk_i^{-1}\right)^{-1}$.
Considering the central distortion constraint, we introduce
\begin{small}\begin{equation}
\underline{k}^{\lozenge}_w=(d_0^{-1}-\sigma_x^{-2})^{-1}. \nonumber
\end{equation}\end{small}\hspace{-1mm}
We now consider two scenarios depending on the relationship
between $\underline{k}^{up}_w$ and $\underline{k}^{\lozenge}_w$:

\textbf{Scenario 1 ($\underline{k}^{\lozenge}_w\geq
\underline{k}^{up}_w$):} The analysis for this scenario is
trivial. It corresponds to the case that the central distortion
constraint is loose. By letting
\begin{small}\begin{equation}
\sigma_j^2=0,\quad j=1,\ldots,J, \nonumber
\end{equation}\end{small}\hspace{-1mm}
the optimal Gaussian test channel is thus specified. The resulting
cental distortion is
\begin{small}\begin{equation}
d_0^{\ast}=\sum_{i=1}^{L} d_l^{-1}-(L-1)\sigma_x^{-2}.
\end{equation}\end{small}\hspace{-1mm}
Further, there is $d_0^{\ast}<=d_0$. In this situation, the
optimal rate for each channel is actually the minimum single
description rate, i.e.
$R_l=\frac{1}{2}\log\frac{\sigma_x^2}{d_l}$, $k=1,\ldots,L$. From
(\ref{equ:lower_bound_loose_central}), it is straightforward that
the Gaussian description scheme achieves the optimal weighted sum
rate.

\textbf{Scenario 2 ($\underline{k}^{\lozenge}_w<
\underline{k}^{up}_w$):} From \emph{Lemma}
\ref{lemma:scalar_monotonicity}, it is ensured that there exists
$\tilde{\sigma}_1^2>0$ such that the corresponding parameter
$\underline{k}_w(\tilde{\sigma}_1^2)$ satisfies
\begin{small}\begin{equation}
\underline{k}_w(\tilde{\sigma}_1^2)=\underline{k}^{\lozenge}_w. \nonumber
\end{equation}\end{small}\hspace{-1mm}
In this situation, there are two possible outcomes when comparing
$\sigma_J^2(\tilde{\sigma}_1^2)$ with $\sigma_x^2$.

We first consider the case that
$\sigma_J^2(\tilde{\sigma}_1^2)<\sigma_x^2$. From \emph{Lemma}
\ref{lemma:scalar_monotonicity}, it follows that
\begin{small}\begin{equation}0<\alpha_1\tilde{\sigma}_1^2<\alpha_2\sigma_2^2(\tilde{\sigma}_1^2)<,\ldots,<\alpha_J^2(\tilde{\sigma}_1^2)<\alpha_J\sigma_x^2.\nonumber\end{equation}\end{small}\hspace{-1mm}
We construct the Gaussian test channel using
$(\tilde{\sigma}_1^2,\ldots,\sigma_J^2(\tilde{\sigma}_1^2))$ and
(\ref{equ:k_l_scalar}). The corresponding multiple description scheme using Gaussian
descriptions satisfies the conditions of \emph{Theorem}
\ref{theorem:optimal_weight_sumRate}. Thus, we obtain the optimal
weighted sum rate.

Next we consider the case that
$\sigma_J^2(\tilde{\sigma}_1^2)\geq\sigma_x^2$. This particular
case actually corresponds to the situation that the individual
distortion constraints associated with the weighting factor
$\alpha_J$ are loose. To show that the outer bound
(\ref{equ:lower_bound}) is still tight, the basic idea is to first
construct a new $d_L'$ smaller than or equal to $d_L$ in
(\ref{equ:k_l_scalar}), and then consider the new multiple description problem with
distortion constraints $(d_1,\ldots,d_{L-1},d_L',d_0)$. The key
observation is that the resulting optimal sum rate
(\ref{equ:lower_bound}) of the new multiple description problem is only a function
of $(d_0,\ldots,d_{L-1},d_0)$, and is unrelated with $d_L'$ or
$d_L$. From \emph{Lemma} \ref{lemma:scalar_monotonicity}, there
exists $0<\bar{\sigma}_1^2 \leq \tilde{\sigma}_1^2$ such that
\begin{small}\begin{equation}
\left\{\begin{array}{l} \sigma_J^2(\bar{\sigma}_1^2) =\sigma_x^2 \\
 \underline{k}^{\lozenge}_w \leq \underline{k}_w(\bar{\sigma}_1^2)\end{array}\right. .\nonumber
\end{equation}\end{small}\hspace{-1mm}
In this situation, we have
\begin{small}\begin{equation}
\left(\underline{k}^{\lozenge}_w+\sigma_J^2(\bar{\sigma}_1^2)\right)^{-1} = \left(\underline{k}_{\{1,\ldots,M_1^{J-1}\}}+\sigma_J^2(\bar{\sigma}_1^2)\right)^{-1}
+\sum_{i=1}^{m_J}\left(k_{M_1^{J-1}+i}+\sigma_J^2(\bar{\sigma}_1^2)\right)^{-1}+\lambda,
\label{equ:A_K_relation_scalar_inequality}
\end{equation}\end{small}\hspace{-1mm}
where $\lambda\geq 0$, and
$\underline{k}_{\{1,\ldots,M_1^{J-1}\}}$ is a function of
$\bar{\sigma}_1^2$ as defined in \emph{Lemma}
\ref{lemma:scalar_monotonicity}. We use the \emph{enhancement}
technique \cite{Weignarten06BC_capacity} to find a $k_L'$ such
that $k_L'\leq k_L$ (correspondingly $d_L'\leq d_L$). Letting
\begin{small}\begin{equation}
\left(k_{L}+\sigma_J^2(\bar{\sigma}_1^2)\right)^{-1}+\lambda= \left(k'_{L}+\sigma_J^2(\bar{\sigma}_1^2)\right)^{-1}\nonumber
\end{equation}\end{small}\hspace{-1mm}
and using the fact that $
\sigma_x^2=\sigma_J^2(\bar{\sigma}_1^2)$, we arrive at
\begin{small}\begin{equation}
k_L'=\left[(k_L+\sigma_x^2)^{-1}+\lambda\right]^{-1}-\sigma_x^2.
\end{equation}\end{small}\hspace{-1mm}
The verification of $k_L'\leq k_L$ is straightforward. Equ.
(\ref{equ:A_K_relation_scalar_inequality}) can be rewritten in
terms of $k_L'$ as
\begin{small}\begin{equation}
\left(\underline{k}^{\lozenge}_0+\sigma_J^2(\bar{\sigma}_1^2)\right)^{-1} = \left(\underline{k}_{\{1,\ldots,M_1^{J-1}\}}+\sigma_J^2(\bar{\sigma}_1^2)\right)^{-1}
+\sum_{i=1}^{m_J-1}\left(k_{M_1^{J-1}+i}+\sigma_J^2(\bar{\sigma}_1^2)\right)^{-1}+\left(k_L'+\sigma_J^2(\bar{\sigma}_1^2)\right)^{-1}.
\end{equation}\end{small}\hspace{-1mm}
From Lemma \ref{lemma:Kw_positive} and Lemma
\ref{lemma:scalar_monotonicity}, we conclude that the Gaussian
test channel built from $(k_1,\ldots,k_{L-1},k_L')$ and
$(\bar{\sigma}_1^2,\ldots,\sigma_J^2(\bar{\sigma}_1^2))$ is valid.
The resulting distortions are $(d_1,\ldots,d_{L-1},d_L',d_0)$,
where $0<d_L' \leq d_L$.

We now show that the optimal weighted sum rate
(\ref{equ:lower_bound}) of the above Gaussian description scheme
is unrelated with $d_L'$. The idea is to construct a sequence
$d_0(\epsilon)$ such that
$\sigma_J^2(\bar{\sigma}_1^2(\epsilon))<\sigma_x^2$, thus allowing
the application of Theorem \ref{theorem:optimal_weight_sumRate}.
Then by letting $\epsilon\rightarrow 0$, the argument is
straightforward. Let
\begin{small}\begin{equation}
d_0(\epsilon)=((\underline{k}^{\diamond}_w+\epsilon)^{-1}+\sigma_x^2)^{-1},
\end{equation}\end{small}\hspace{-1mm}
where $\epsilon$ is chosen such that $\epsilon>0$ and
$(\underline{k}^{\diamond}_w+\epsilon)^{-1}>\sum_{i=1}^{L-1}k_i^{-1}+k_L'^{-1}$.
Taking $(k_1,\ldots,k_{L-1},k_L')$ in Lemma
\ref{lemma:scalar_monotonicity}, we obtain
\begin{small}\begin{eqnarray}
&&\underline{k}^{\diamond}_w+\epsilon=\underline{k}_w(\bar{\sigma}_1^2(\epsilon)), \nonumber\\
&& \sigma_J^2(\bar{\sigma}_1^2(\epsilon))<\sigma_x^2, \quad \epsilon \neq 0. \nonumber
\end{eqnarray}\end{small}\hspace{-1mm}
Considering the multiple description problem with distortion constraints
($d_1,\ldots,d_L',d_0(\epsilon)$), the situation corresponds to
the one discussed in the first case. As long as $\epsilon \neq 0$,
the optimal weighted sum rate of the modified distortion multiple description
problem is known, which is given by (\ref{equ:lower_bound}).
Particularly, the expressions of the optimal $n_j(\epsilon)$
($n_j$ is the scalar version of $\boldsymbol{N}_j$ in
(\ref{equ:lower_bound})), $j=1,\ldots,J$ take the form (see
Proposition \ref{proposition:optimal_weighted_sumrate})
\begin{small}\begin{equation}
n_j(\epsilon)=\frac{\sigma_x^2}{\sigma_x^2/\sigma_j^2(\bar{\sigma}_1^2(\epsilon))-1}, \quad j=1,\ldots,J.
\label{equ:n_j_sigma_j_relation_limit}
\end{equation}\end{small}\hspace{-1mm}
Taking the limit of $\epsilon$, i.e. $\epsilon\downarrow 0$, there
are $d_0(\epsilon)\downarrow d_0$,
$\sigma_J^2(\bar{\sigma}_1^2(\epsilon))\uparrow \sigma_x^2$ and
$n_J(\epsilon)\uparrow +\infty$. Due to the limiting operation
$n_J(\epsilon)\uparrow +\infty$, the side distortion $d_L'$ does
not contribute to the resulting optimal weighted sum rate
(\ref{equ:lower_bound}). In other words, the optimal weighted sum
rate (\ref{equ:lower_bound}) is only a function of
$(d_1,\ldots,k_{L-1},d_0)$. Thus, we show that the outer bound
(\ref{equ:lower_bound}) is still tight for the original multiple description problem
with distortion constraints $(d_1,\ldots,d_L,d_0)$ (i.e., the
individual side distortions are loose) .

In summary, for a scalar Gaussian source, the Gaussian description
scheme fully characterizes the rate region. When the central
distortion constraint is trivial ($\underline{k}^{\lozenge}_w\geq
\underline{k}^{up}_w$), the shape of the rate region belongs to
the class of contra-polymatroids \cite{Welsh76Matroid}. If it is
not the case, one can always construct an optimal Gaussian test
channel satisfying the distortion constraint. This is to say that
when $\underline{k}^{\lozenge}_w< \underline{k}^{up}_w$, there
always exists
$0<\alpha_1\sigma_1^2<,\ldots,<\alpha_J\sigma_J^2\leq
\alpha_J\sigma_x^2$ such that
\begin{small}\begin{eqnarray}
&& \alpha_1\sum_{i=1}^{m_1}(\sigma_1^2+k_i)^{-1}=\frac{\alpha_2}{\sigma_1^2}-\frac{\alpha_1-\alpha_2}{\sigma_2^2-\sigma_1^2}, \label{equ:KKT_scalar_1}\\
&& \alpha_j\sum_{i=1}^{m_j}(\sigma_j^2+k_{M_1^{j-1}+i})^{-1}+ \frac{\alpha_{j-1}}{\sigma_j^2}- \frac{\alpha_{j-1}-\alpha_j}{\sigma_j^2-\sigma_{j-1}^2}
= \frac{\alpha_{j+1}}{\sigma_j^2}-\frac{\alpha_j-\alpha_{j+1}}{\sigma_{j+1}^2-\sigma_j^2}, \quad j=2,\ldots,J-1, \label{equ:KKT_scalar_2}\\
&& \lambda+\alpha_J\sum_{i=1}^{m_J}(\sigma_J^2+k_{M_1^{J-1}+i})^{-1}+ \frac{\alpha_{J-1}}{\sigma_J^2}- \frac{\alpha_{J-1}-\alpha_J}{\sigma_J^2-\sigma_{J-1}^2}
= \alpha_J(\sigma_J^2+k^{\lozenge}_0)^{-1}, \label{equ:KKT_scalar_3}
\end{eqnarray}\end{small}\hspace{-1mm}
where
\begin{small}\begin{equation}
\lambda(\sigma_x^2-\sigma_J^2)=0, \quad \lambda\geq 0, \label{equ:KKT_scalar_4}
\end{equation}\end{small}\hspace{-1mm}
and $k_l$, $l=1,\ldots,L$ are as defined in
(\ref{equ:k_l_scalar}).  The conditions
(\ref{equ:KKT_scalar_1})-(\ref{equ:KKT_scalar_4}) follow from
(\ref{equ:A_K_relation_scalar})-(\ref{equ:A_K_relation_scalar2}),
(\ref{equ:k_l_scalar}) and
(\ref{equ:A_K_relation_scalar_inequality}). Further, the solution
to (\ref{equ:KKT_scalar_1})-(\ref{equ:KKT_scalar_4}) is unique.

The solution to (\ref{equ:KKT_scalar_1})-(\ref{equ:KKT_scalar_4})
can be interpreted as a solution to an optimization problem. We
first introduce parameters $y_j>0$, $j=1,\ldots,J$ where
$\sigma_j^2=\sum_{i=1}^j y_i$. Define a new function
$F(y_1,\ldots,y_J)$ as
\begin{small}\begin{eqnarray}
&&F(y_1,\ldots,y_J)=\alpha_J\log\left(\underline{k}^{\lozenge}_w+\sum_{d=1}^{J}y_d\right)-\sum_{j=1}^{J}\alpha_j\sum_{i=1}^{m_j}\log\left(\sum_{d=1}^j
y_d+k_{M_1^{j-1}+i}\right)
+\sum_{j=1}^{J-1}(\alpha_j-\alpha_{j+1})\log y_{j+1}\nonumber\\
&&-\sum_{j=2}^{J-1}(\alpha_{j-1}-\alpha_{j+1})\log\left(\sum_{i=1}^j
y_i\right)+\alpha_2\log y_1-\alpha_{J-1}\log(\sum_{i=1}^J y_i).
\end{eqnarray}\end{small}\hspace{-1mm}
Consider the following optimization problem:
\begin{small}\begin{equation}
\max_{\{y_j\}_{j=1}^{J}} F(y_1,\ldots,y_J)\quad \textrm{ subject to } \left\{\begin{array}{l}
0<y_j,\quad j=1,\ldots,J \\
\sum_{j=1}^J y_j\leq \sigma_x^2
\end{array} \right..
\end{equation}\end{small}\hspace{-1mm}
Using the concept of \emph{Lagrange duality} \cite[Chapter
5]{Boyd04ConvexOptimization} to handle the constraint
$\sum_{j=1}^J y_j\leq \sigma_x^2$, the corresponding
Karush-Kuhn-Tucker (KKT) conditions take the form
\begin{small}\begin{eqnarray}
&&\frac{\partial F}{\partial y_j}-\gamma=0, \quad j=1,\ldots,J\label{equ:KKT_scalar_a}\\
&&\gamma(\sigma_x^2-\sum_{j=1}^J y_j)=0,  \label{equ:KKT_scalar_b}\\
&& \sum_{j=1}^J y_j\leq \sigma_x^2 \textrm{ and } y_j>0, j=1,\ldots,J. \label{equ:KKT_scalar_c}
\end{eqnarray}\end{small}\hspace{-1mm}
One can show that
(\ref{equ:KKT_scalar_a})-(\ref{equ:KKT_scalar_c}) are equivalent
to (\ref{equ:KKT_scalar_1})-(\ref{equ:KKT_scalar_4}) by plugging
$y_j=\sigma_j^2-\sigma_{j-1}^2$ and $\gamma=\lambda$ into the
expressions. Thus, instead of solving a set of equations, we can
equivalently consider an optimization problem. The sufficient
condition making the KKT conditions hold is
$\underline{k}^{\lozenge}_w< \underline{k}^{up}_w$ as discussed in
\textbf{Scenario 2}.


\section{Conclusion}
We have addressed the rate region of the multiple description problem with respect
to individual and central distortion constraints for a vector
Gaussian source. An outer bound was derived for the considered
rate region. In the special case of a scalar Gaussian source, the
lower bound was shown to be tight.

The work in \cite{Wang07VectorMDC} treated the special case that
all the weighting factors in
(\ref{equ:weighted_sume_optimization}) are identical (i.e.,
$J=1$), which corresponds to minimizing the sum rate. In this
particular case, it was previously shown that the outer bound
given by (\ref{equ:lower_bound_sumRate}), remains tight even if
some of the distortion constraints are loose. Thus, the optimal
sum rate for individual and central receivers has been completely
characterized \cite{Wang07VectorMDC}. While we have not been able
to prove similar result for the weighted sum rate case (where
$J>1$), we believe that the outer bound given by
(\ref{equ:lower_bound}), remains tight even if some of the
distortion constraints are loose.


%

\appendices

\section{Useful Matrix Lemmas}
\begin{lemma}[Matrix Inversion Lemma] \cite[Theorem 2.5]{Zhang99MatrixTheory}
Let $\boldsymbol{A}$ be an $m\times m$ nonsingular matrix and
$\boldsymbol{B}$ be an $n\times n$ nonsingular matrix and let
$\boldsymbol{C}$ and $\boldsymbol{D}$ be $m\times n$ and $n\times
n$ matrices, respectively. If the matrix
$\boldsymbol{A}+\boldsymbol{CBD}$ is nonsingular, then
\begin{small}\begin{equation}
(\boldsymbol{A}+\boldsymbol{CBD})^{-1}=\boldsymbol{A}^{-1}-\boldsymbol{A}^{-1}\boldsymbol{C}(\boldsymbol{B}^{-1}+\boldsymbol{D}\boldsymbol{A}^{-1}\boldsymbol{C})^{-1}\boldsymbol{DA}^{-1}.
\end{equation}\end{small}\hspace{-1mm}
\label{lemma:matrix_inversion}
\end{lemma}

\begin{lemma} \cite[Theorems 6.8 and 6.9]{Zhang99MatrixTheory}
Let $\boldsymbol{A}$ and $\boldsymbol{B}$ be positive definite
matrices such that $\boldsymbol{A}\succ\boldsymbol{B}$
$(\boldsymbol{A}\succeq \boldsymbol{B})$. Then
\begin{small}\begin{equation}
\hspace{-7mm}|\boldsymbol{A}|\succ|\boldsymbol{B}|\textrm{ }
(|\boldsymbol{A}|\succeq|\boldsymbol{B}|) \nonumber
\end{equation}
\begin{equation}
\boldsymbol{A}^{-1} \prec \boldsymbol{B}^{-1} \textrm{ }
(\boldsymbol{A}^{-1} \preceq \boldsymbol{B}^{-1}).
\end{equation}\end{small}\hspace{-1mm}
\label{lemma:ordering_relation}
\end{lemma}

\section{Proof of Lemma \ref{lemma:EPI}}
\label{appendix:Proof of Lemma:EPI}

The main tool used in the proof is the generalized Costa's
entropy-power inequality (EPI) developed by R. Liu et al. in
\cite{RuohengLiu09VectorEPI}. Specifically, we use the conditional
version of the generalized Costa's EPI, which was shown to be a
simple extension of the generalized Costa's EPI
\cite{RuohengLiu09VectorEPI}. To make the work complete, we
present the result in a lemma below.

\begin{lemma}[Theorem 1, Corollary 1 in \cite{RuohengLiu09VectorEPI}]
Let $\boldsymbol{z}$ be a Gaussian random n-vector with a positive
definite covariance matrix $\boldsymbol{C}$, and let
$\boldsymbol{A}$ be an $n\times n$ real symmetric matrix such that
$\boldsymbol{0}\preceq \boldsymbol{A} \preceq \boldsymbol{I}_n$.
Then
\begin{small}\begin{equation}
\exp\left[\frac{2}{n}h(\boldsymbol{p}+\boldsymbol{A}^{\frac{1}{2}}\boldsymbol{z}|\boldsymbol{v})\right]
\geq |\boldsymbol{I}-\boldsymbol{A}|^{\frac{1}{n}}\exp\left[\frac{2}{n}h(\boldsymbol{p}|\boldsymbol{v})\right]
+|\boldsymbol{A}|^{\frac{1}{n}}\exp\left[\frac{2}{n}h(\boldsymbol{p}+\boldsymbol{z}|\boldsymbol{v})\right]
\label{equ:cost_EPI}
\end{equation}\end{small}\hspace{-1mm}
for any $\boldsymbol{v}$ and $n$-vector $\boldsymbol{p}$
independent of $\boldsymbol{z}$. The equality holds if
$(\boldsymbol{p},\boldsymbol{v})$ are jointly Gaussian with a
conditional covariance matrix
$\boldsymbol{B}=\textrm{Cov}[\boldsymbol{p}|\boldsymbol{v}]$ such
that $\boldsymbol{B}-\boldsymbol{AB}$ and
$\boldsymbol{B}+\boldsymbol{A}^{\frac{1}{2}}\boldsymbol{C}\boldsymbol{A}^{\frac{1}{2}}$
are proportional. \label{lemma:Costa_EPI}
\end{lemma}

Next we are in a position to prove the lemma. The basic idea is as
follows. we first apply a linear transformation to the
$nN$-dimensional random vectors $\boldsymbol{x}^n$,
$\boldsymbol{z}_1^n$ and $\boldsymbol{z}_2^n$ such that after
transformation the random vectors corresponding to
$\boldsymbol{z}_1^n$ and $\boldsymbol{z}_2^n$ have diagonal
covariance matrices. The Costa's EPI (\ref{equ:cost_EPI}) is then
used to prove an extremal entropy inequality in the transform
domain. Finally, the upper-bound inequality
(\ref{equ:extreme_inequality}) is obtained by converting the
extremal entropy inequality to the original domain.

We consider diagonalizing $\boldsymbol{N}_1$ and
$\boldsymbol{N}_2$ simultaneously.  Using the fact that
$\boldsymbol{N}_1$ and $\boldsymbol{N}_2$ are positive definite,
it is known \cite{Strang98Algebra} that there exists an invertible
matrix $\boldsymbol{U}$ such that
\begin{small}\begin{equation}\boldsymbol{U}^{t}\boldsymbol{N_1}\boldsymbol{U}=\boldsymbol{\Lambda}_1
\textrm{ and }
\boldsymbol{U}^{t}\boldsymbol{N_2}\boldsymbol{U}=\boldsymbol{\Lambda}_2,\end{equation}\end{small}\hspace{-1mm}
where $\boldsymbol{\Lambda}_1$ and $\boldsymbol{\Lambda}_2$ are
positive definite diagonal matrices. From the assumption on
$\boldsymbol{N}_1$ and $\boldsymbol{N}_2$, it is immediate that
$\boldsymbol{\Lambda}_1\prec \boldsymbol{\Lambda}_2$. Let
\begin{small}\begin{eqnarray}
&&\boldsymbol{y}_1^{n}=(\boldsymbol{I}_n\otimes\boldsymbol{U}^t) (\boldsymbol{x}^{n}+\boldsymbol{z}_1^{n}) \label{equ:y_1}\\
&&\hspace{-10mm}\textrm{ and}\quad \boldsymbol{y}_2^{n}=(\boldsymbol{I}_n\otimes \boldsymbol{U}^t) (\boldsymbol{x}^{n}+\boldsymbol{z}_2^{n})\label{equ:y_2}.
\end{eqnarray}\end{small}\hspace{-1mm}
Note that by applying the linear transformation in
(\ref{equ:y_1})-(\ref{equ:y_2}), the resulting random vectors
corresponding to $\boldsymbol{z}_1^{n}$ and $\boldsymbol{z}_2^{n}$
have independent components. To simplify the notation in the
derivation, we introduce
\begin{small}\begin{eqnarray}
&&\tilde{\boldsymbol{x}}^{n}= (\boldsymbol{I}_n\otimes\boldsymbol{U}^t )\boldsymbol{x}^{n} \label{equ:x_transform}\\
&&\hspace{-10mm}\textrm{ and}\quad \tilde{\boldsymbol{z}}^{n}= (\boldsymbol{I}_n\otimes\boldsymbol{U}^t )\boldsymbol{z}_2^{n}.
\end{eqnarray}\end{small}\hspace{-1mm}
Consequently we have
$\textrm{Cov}[\tilde{\boldsymbol{z}}^{n}]=\boldsymbol{I}_n\otimes\boldsymbol{\Lambda}_2
$. The two variables $\boldsymbol{y}_1^{n}$ and
$\boldsymbol{y}_2^{n}$ can be equivalently written as
\begin{small}\begin{eqnarray}
&&\boldsymbol{y}_2^{n}= \tilde{\boldsymbol{x}}^{n}+\tilde{\boldsymbol{z}}^{n}\\
&&\boldsymbol{y}_1^{n}= \tilde{\boldsymbol{x}}^{n}+(\boldsymbol{I}_n\otimes\boldsymbol{A}^{\frac{1}{2}} )\tilde{\boldsymbol{z}}^{n},
\end{eqnarray}\end{small}\hspace{-1mm} where $\boldsymbol{A}=\boldsymbol{\Lambda}_1\boldsymbol{\Lambda}_2^{-1}$ satisfies $\boldsymbol{0}\prec\boldsymbol{A}\prec \boldsymbol{I}$.
Considering the quantity $h(\boldsymbol{y}_1^{n}|\boldsymbol{v})$,
it is immediate from Lemma \ref{lemma:Costa_EPI} that
\begin{small}\begin{equation}
h(\boldsymbol{y}_1^{n}|\boldsymbol{v})\geq
\frac{Nn}{2}\log\left[|\boldsymbol{I}_{N}-\boldsymbol{A}|^{\frac{1}{N}}\exp\left[\frac{2}{Nn}h(\tilde{\boldsymbol{x}}^{n}|\boldsymbol{v})\right]
+|\boldsymbol{A}|^{\frac{1}{N}}\exp\left[\frac{2}{Nn}h(\boldsymbol{y}_2^{n}|\boldsymbol{v})\right]\right].
\label{equ:Costa_EPI_apply}
\end{equation}\end{small}\hspace{-1mm}

We now prove an extremal entropy inequality in the transform
domain. By using (\ref{equ:Costa_EPI_apply}), we have
\begin{small}\begin{eqnarray}
&&\hspace{-3mm}\mu_2h(\boldsymbol{y}_2^{n}|\boldsymbol{v})-\mu_1
h(\boldsymbol{y}_1^{n}|\boldsymbol{v})
+(\mu_1-\mu_2)h(\tilde{\boldsymbol{x}}^{n}|\boldsymbol{v}) \nonumber\\
&&\hspace{-3mm}\leq  \mu_2h(\boldsymbol{y}_2^{n}|\boldsymbol{v})-
\frac{\mu_1
Nn}{2}\log\left[|\boldsymbol{I}_{N}-\boldsymbol{A}|^{\frac{1}{N}}\exp\left[\frac{2}{Nn}h(\tilde{\boldsymbol{x}}^{n}|\boldsymbol{v})\right]
+|\boldsymbol{A}|^{\frac{1}{N}}\exp\left[\frac{2}{Nn}h(\boldsymbol{y}_2^{n}|\boldsymbol{v})\right]\right]
+(\mu_1-\mu_2)h(\tilde{\boldsymbol{x}}^{n}|\boldsymbol{v}) \nonumber \\
&&\hspace{-3mm}
=\mu_2I(\tilde{\boldsymbol{z}}^{n};\tilde{\boldsymbol{x}}^{n}+\tilde{\boldsymbol{z}}^{n}|\boldsymbol{v})
-\frac{\mu_1Nn}{2}\log\left[
|\boldsymbol{I}_N-\boldsymbol{A}|^{\frac{1}{N}}+|\boldsymbol{A}|^{\frac{1}{N}}\exp\left[\frac{2}{Nn}I(\tilde{\boldsymbol{z}}^{n};\tilde{\boldsymbol{x}}^{n}+\tilde{\boldsymbol{z}}^{n}|\boldsymbol{v})\right]
\right]. \label{equ:Costa_EPI_apply_2}
\end{eqnarray}\end{small}\hspace{-1mm}
From (\ref{equ:Costa_EPI_apply_2}), the function
\begin{small}\begin{equation}
f(t)=\mu_2t-\frac{\mu_1Nn}{2}\log\left(|\boldsymbol{I}_N-\boldsymbol{A}|^{\frac{1}{N}}+|\boldsymbol{A}|^{\frac{1}{N}}\exp\left[\frac{2}{Nn}t\right]\right)
\end{equation}\end{small}\hspace{-1mm}
is concave in t and has a global maxima at
\begin{small}\begin{equation}
t^{\ast}=\frac{Nn}{2}\log\left(\frac{\mu_2|\boldsymbol{A}^{-1}-\boldsymbol{I}_N|^{\frac{1}{N}}}{\mu_1-\mu_2}\right).
\label{equ:t_optimal}
\end{equation}\end{small}\hspace{-1mm}
Combining (\ref{equ:Costa_EPI_apply_2}) and (\ref{equ:t_optimal})
produces
\begin{small}\begin{equation}
\mu_2h(\boldsymbol{y}_2^{n}|\boldsymbol{v})-\mu_1 h(\boldsymbol{y}_1^{n}|\boldsymbol{v})
+(\mu_1-\mu_2)h(\tilde{\boldsymbol{x}}^{n}|\boldsymbol{v}) \leq \frac{\mu_2 n}{2}\log\frac{\mu_2^N|\boldsymbol{\Lambda}_2-
\boldsymbol{\Lambda}_1|}{(\mu_1-\mu_2)^N|\boldsymbol{\Lambda}_1|}-\frac{\mu_1 n}{2}\log\frac{\mu_1^N|\boldsymbol{\Lambda}_2-\boldsymbol{\Lambda}_1|}{(\mu_1-\mu_2)^N|\boldsymbol{\Lambda}_2|}.
\label{equ:extreme_inequality_2}
\end{equation}\end{small}\hspace{-1mm}
The equality in (\ref{equ:extreme_inequality_2}) holds if
$(\tilde{\boldsymbol{x}}^{n},\boldsymbol{v})$ satisfies the
equality condition in (\ref{equ:Costa_EPI_apply}) imposed by the
generalized Costa's EPI and the optimality condition
(\ref{equ:t_optimal}). From Lemma \ref{lemma:Costa_EPI}, the
equality condition to (\ref{equ:Costa_EPI_apply}) can be
mathematically written as
\begin{small}\begin{equation}
\left(\boldsymbol{I}_{nN}-\boldsymbol{I}_n\otimes \boldsymbol{A}\right)\textrm{Cov}(\tilde{\boldsymbol{x}}^n|\boldsymbol{v})
=c\left[\textrm{Cov}(\tilde{\boldsymbol{x}}^n|\boldsymbol{v})+\boldsymbol{I}_n\otimes (\boldsymbol{A}^{1/2}\boldsymbol{\Lambda}_2\boldsymbol{A}^{1/2})\right],
\label{equ:epi_proportionality}
\end{equation}\end{small}\hspace{-1mm}
where $c$ is a scalar variable. By combining (\ref{equ:t_optimal})
and (\ref{equ:epi_proportionality}), it is found that the above
two conditions are equivalent to the fact that
$(\tilde{\boldsymbol{x}}^{n},\boldsymbol{v})$ are jointly Gaussian
with a conditional covariance matrix
\begin{small}\begin{equation}
\textrm{Cov}[\tilde{\boldsymbol{x}}^{n}|\boldsymbol{v}]= \boldsymbol{I}_n\otimes\left[(\mu_1-\mu_2)\boldsymbol{\Lambda}_2(\mu_2\boldsymbol{\Lambda}_2-\mu_1\boldsymbol{\Lambda}_1)^{-1}\boldsymbol{\Lambda}_1\right].
\label{equ:covariance_condition_2}
\end{equation}\end{small}\hspace{-1mm}

The final step is to convert the inequality
(\ref{equ:extreme_inequality_2}) to an inequality in the original
domain. In other words, we relate
(\ref{equ:extreme_inequality_2})-(\ref{equ:covariance_condition_2})
to
(\ref{equ:extreme_inequality})-(\ref{equ:covariance_condition}).
From (\ref{equ:y_1})-(\ref{equ:x_transform}), we arrive at
\begin{small}\begin{eqnarray}
&&h(\boldsymbol{y}_i^{n}|\boldsymbol{v})=h(\boldsymbol{x}^{n}+\boldsymbol{z}_i^{n}|\boldsymbol{v})+n\log|\boldsymbol{U}^t|, \quad i=1,2 \label{equ:trans_relation_1}\\
&&h(\tilde{\boldsymbol{x}}^{n}|\boldsymbol{v})=h(\boldsymbol{x}^{n}|\boldsymbol{v})+n\log|\boldsymbol{U}^t| \label{equ:trans_relation_2} \\
&&\textrm{Cov}[\tilde{\boldsymbol{x}}^{n}|\boldsymbol{v}]=\left[\boldsymbol{I}_n\otimes\boldsymbol{U}^t
\right]\textrm{Cov}[\boldsymbol{x}^{n}|\boldsymbol{v}]\left[\boldsymbol{I}_n\otimes\boldsymbol{U}
\right]
.\label{equ:trans_relation_3}\end{eqnarray}\end{small}\hspace{-1mm}
By plugging
(\ref{equ:trans_relation_1})-(\ref{equ:trans_relation_3}) into
(\ref{equ:extreme_inequality_2})-(\ref{equ:covariance_condition_2}),
the derivation of
(\ref{equ:extreme_inequality})-(\ref{equ:covariance_condition}) is
then straightforward. The condition $\boldsymbol{0}\prec
\mu_1\boldsymbol{N}_1\prec \mu_2\boldsymbol{N}_2$ imposed on
$\boldsymbol{N}_1$ and $\boldsymbol{N}_2$ is to ensure that the
conditional covariance matrix
$\textrm{Cov}[\boldsymbol{x}^n|\boldsymbol{v}]$ in
(\ref{equ:covariance_condition}) is positive definite.

\section{Proof of Lemma \ref{lemma:A_K_relation}}
\label{appendix:proof_lemma_A_K_relation}

The proof is essentially the same for every $j=1,\ldots,J$. Thus
we only consider the derivation for a particular $j$. we first let
$\boldsymbol{A}_j\succ \boldsymbol{0}$. By using Lemma
\ref{lemma:matrix_inversion}, there is
\begin{small}\begin{eqnarray}
&&\hspace{-7mm}\left[\boldsymbol{A}_j^{-1}+(\boldsymbol{I}_N,\ldots,\boldsymbol{I}_N)\boldsymbol{K}_{\{1,\ldots,M_1^j\}}^{-1}(\boldsymbol{I}_N,\ldots,\boldsymbol{I}_N)^t\right]^{-1} \nonumber\\
&=& \boldsymbol{A}_j-\boldsymbol{A}_j(\boldsymbol{I}_N,\ldots,\boldsymbol{I}_N)\left[\boldsymbol{K}_{\{1,\ldots,M_1^j\}}
+(\boldsymbol{I}_N,\ldots,\boldsymbol{I}_N)^t\boldsymbol{A}_j(\boldsymbol{I}_N,\ldots,\boldsymbol{I}_N)\right]^{-1}(\boldsymbol{I}_N,\ldots,\boldsymbol{I}_N)^t\boldsymbol{A}_j \nonumber \\
&=&
\boldsymbol{A}_j-\boldsymbol{A}_j\Bigg[(\boldsymbol{I}_N,\ldots,\boldsymbol{I}_N)\left(\boldsymbol{K}_{\{1,\ldots,M_1^{j-1}\}}
+\boldsymbol{H}\otimes
\boldsymbol{A}_j\right)^{-1}(\boldsymbol{I}_N,\ldots,\boldsymbol{I}_N)^t
\nonumber\\&&+\sum_{i=1}^{m_j}\left(\boldsymbol{K}_{M_1^{j-1}+i}+\boldsymbol{A}_j\right)^{-1}\Bigg]\boldsymbol{A}_j.
\label{equ:proof_A_K_relation}
\end{eqnarray}\end{small}\hspace{-1mm}
By rearranging the items in (\ref{equ:proof_A_K_relation}), we
have
\begin{small}\begin{eqnarray}
&&\hspace{-7mm}(\boldsymbol{I}_N,\ldots,\boldsymbol{I}_N)\boldsymbol{K}_{\{1,\ldots,M_1^j\}}^{-1}(\boldsymbol{I}_N,\ldots,\boldsymbol{I}_N)^t\nonumber \\
&=&\hspace{-2mm}\Bigg[\boldsymbol{A}_j-\boldsymbol{A}_j\Bigg((\boldsymbol{I}_N,\ldots,\boldsymbol{I}_N)\left(\boldsymbol{K}_{\{1,\ldots,M_1^{j-1}\}}
+\boldsymbol{H}\otimes
\boldsymbol{A}_j\right)^{-1}(\boldsymbol{I}_N,\ldots,\boldsymbol{I}_N)^t+
\nonumber \\
&&\hspace{-2mm}\sum_{i=1}^{m_j}\left(\boldsymbol{K}_{M_1^{j-1}+i}+\boldsymbol{A}_j\right)^{-1}\Bigg)\boldsymbol{A}_j\Bigg]^{-1}-\boldsymbol{A}_j^{-1} \nonumber \\
&=&
\hspace{-2mm}\Bigg[-\boldsymbol{A}_j+\Bigg((\boldsymbol{I}_N,\ldots,\boldsymbol{I}_N)\left(\boldsymbol{K}_{\{1,\ldots,M_1^{j-1}\}}
+\boldsymbol{H}\otimes
\boldsymbol{A}_j\right)^{-1}(\boldsymbol{I}_N,\ldots,\boldsymbol{I}_N)^t+
\nonumber \\
&&\hspace{-2mm}\sum_{i=1}^{m_j}\left(\boldsymbol{K}_{M_1^{j-1}+i}+\boldsymbol{A}_j\right)^{-1}\Bigg)^{-1}\Bigg]^{-1}.\label{equ:proof_A_K_relation_1}
\end{eqnarray}\end{small}\hspace{-1mm}
With the definition of $\underline{\boldsymbol{K}}_S$ in
(\ref{equ:virtual_channel_matrix}),
(\ref{equ:proof_A_K_relation_1}) can be rewritten as
\begin{small}\begin{equation}
\left(\underline{\boldsymbol{K}}_{\{1,\ldots,M_1^j\}}+\boldsymbol{A}_j\right)^{-1}\hspace{-0.5mm}=\hspace{-0.5mm}(\boldsymbol{I}_N,\ldots,\boldsymbol{I}_N)\left(\boldsymbol{K}_{\{1,\ldots,M_1^{j-1}\}}
+\boldsymbol{H}\otimes
\boldsymbol{A}_j\right)^{-1}\hspace{-0.5mm}(\boldsymbol{I}_N,\ldots,\boldsymbol{I}_N)^t
+\sum_{i=1}^{m_j}\left(\boldsymbol{K}_{M_1^{j-1}+i}+\boldsymbol{A}_j\right)^{-1}.
\label{equ:proof_A_K_relation_2}
\end{equation}\end{small}\hspace{-1mm}

We now consider simplifying (\ref{equ:proof_A_K_relation_2})
further. By using Lemma  \ref{lemma:matrix_inversion}, there is
\begin{small}\begin{eqnarray}
&&\hspace{-6mm}\left[(-\boldsymbol{A}_j)^{-1}+(\boldsymbol{I}_N,\ldots,\boldsymbol{I}_N)\left(\boldsymbol{K}_{\{1,\ldots,M_1^{j-1}\}}
+\boldsymbol{H}\otimes \boldsymbol{A}_j\right)^{-1}(\boldsymbol{I}_N,\ldots,\boldsymbol{I}_N)^t\right]^{-1} \nonumber \\
&=& -\boldsymbol{A}_j-(-\boldsymbol{A}_j)\left(\boldsymbol{I}_N,\ldots,\boldsymbol{I}_N\right)
\boldsymbol{K}_{\{1,\ldots,M_1^{j-1}\}}^{-1}
(\boldsymbol{I}_N,\ldots,\boldsymbol{I}_N)^t(-\boldsymbol{A}_j). \label{equ:proof_A_K_relation_2.5}
\end{eqnarray}\end{small}\hspace{-1mm}
Similarly to the derivation of (\ref{equ:proof_A_K_relation_1})
from (\ref{equ:proof_A_K_relation}),
(\ref{equ:proof_A_K_relation_2.5}) can be rewritten as
\begin{small}\begin{eqnarray}
\hspace{-5mm}&&(\boldsymbol{I}_N,\ldots,\boldsymbol{I}_N)\left(\boldsymbol{K}_{\{1,\ldots,M_1^{j-1}\}}
+\boldsymbol{H}\otimes \boldsymbol{A}_j\right)^{-1}(\boldsymbol{I}_N,\ldots,\boldsymbol{I}_N)^t \nonumber \\
&=& \left[-\boldsymbol{A}_j-\boldsymbol{A}_j\left(\boldsymbol{I}_N,\ldots,\boldsymbol{I}_N\right)
\boldsymbol{K}_{\{1,\ldots,M_1^{j-1}\}}^{-1}
(\boldsymbol{I}_N,\ldots,\boldsymbol{I}_N)^t\boldsymbol{A}_j\right]^{-1}+\boldsymbol{A}_j^{-1} \nonumber\\
&=& \left[ \left(\left(\boldsymbol{I}_N,\ldots,\boldsymbol{I}_N\right)
\boldsymbol{K}_{\{1,\ldots,M_1^{j-1}\}}^{-1}
(\boldsymbol{I}_N,\ldots,\boldsymbol{I}_N)^t\right)^{-1}+\boldsymbol{A}_j\right]^{-1} \nonumber \\
&=&\left(\underline{\boldsymbol{K}}_{\{1,\ldots,M_1^{j-1}\}}+\boldsymbol{A}_j\right)^{-1} \label{equ:proof_A_K_relation_3}.
\end{eqnarray}\end{small}\hspace{-1mm}
Combining (\ref{equ:proof_A_K_relation_2}) and
(\ref{equ:proof_A_K_relation_3}) produces
(\ref{equ:A_K_relation}). For the case that
$\boldsymbol{A}_j\succeq \boldsymbol{0}$, there exists $\delta>0$
such that
$\boldsymbol{A}_j+\epsilon\boldsymbol{I}_N\succ\boldsymbol{0}$ and
the corresponding $K_{\{1,\ldots,M_1^j\}}\succ\boldsymbol{0}$ for
 $\epsilon\in(0,\delta)$, which supports the previous argument.
By letting $\epsilon\rightarrow 0^{+}$, we obtain the result. The
proof is complete.

\section{Proof of Lemma \ref{lemma:scalar_monotonicity}}
\label{appendix:proof_of_scalar_monotonicity} We prove the lemma
using induction argument. We first consider the monotonicity
properties of $\sigma_2^2$ and
$\underline{k}_{\{1,\ldots,M_1^1\}}$ over $\sigma_1^2$, referred
to as \textbf{Case 1}. Later, we extend the analysis to the
general case. The proof is rather long. We present the proof in
several steps.

\textbf{Case 1 ($j=1$)}: In order to discuss the properties of
$\sigma_2^2$ and $\underline{k}_{\{1,\ldots,M_1^1\}}$ as functions
of $\sigma_1^2$, we first study the support region of $\sigma_1^2$
such that $\sigma_2^2 > \sigma_1^2$. By combining
(\ref{equ:A_K_relation_scalar}) and
(\ref{equ:A_K_relation_scalar2}), we obtain
\begin{small}\begin{equation}
\frac{\alpha_2}{\alpha_1\sigma_1^{2}}+\frac{\alpha_2-\alpha_1}{\alpha_1(\sigma_2^2-\sigma_1^2)}=\sum_{i=1}^{m_1}\left(\sigma_1^2+k_{i}\right)^{-1}. \nonumber
\end{equation}\end{small}\hspace{-1mm}
In order that $\sigma_2^2>\sigma_1^2$, there is
\begin{small}\begin{eqnarray}
&&\frac{\alpha_2}{\alpha_1\sigma_1^{2}}>\sum_{i=1}^{m_1}\left(\sigma_1^2+k_{i}\right)^{-1}\stackrel{\sigma_1^2>0}{\Rightarrow}\sum_{i=1}^{m_1}\frac{k_i}{\sigma_1^2+k_i}>m_1-\frac{\alpha_2}{\alpha_1}.
\nonumber\end{eqnarray}\end{small}\hspace{-1mm} Letting
\begin{small}\begin{equation}f(\sigma_1^2)=\sum_{i=1}^{m_1}\frac{k_i}{\sigma_1^2+k_i}-\left(m_1-\frac{\alpha_2}{\alpha_1}\right),\end{equation}\end{small}\hspace{-1mm}
we have  $f(0)>0$ and $f(+\infty)<0$.  By using intermediate value
theorem \cite[p. 48]{Royden88RealAnalysis}, it is obvious that
there exists $\hat{\sigma}_{\{1\}}^2>0$ such that
\begin{small}\begin{eqnarray}
&& 0<\sigma_1^2<\hat{\sigma}_{\{1\}}^2\Rightarrow \sigma_2^2>\sigma_1^2 \label{equ:mono_proof1}\\
\textrm{and}&& \left\{\begin{array}{l}\sigma_2^2\rightarrow \sigma_1^2 \textrm{ as }  \sigma_1^2\rightarrow 0 \\
 \sigma_2^2\rightarrow +\infty \textrm{ as }  \sigma_1^2\rightarrow \hat{\sigma}_{\{1\}}^2
 \end{array}.\right. \label{equ:mono_proof2}
\end{eqnarray}\end{small}\hspace{-1mm}
Further, by analyzing
(\ref{equ:A_K_relation_scalar})-(\ref{equ:A_K_relation_scalar2})
under the situation
(\ref{equ:mono_proof1})-(\ref{equ:mono_proof2}), we find that
$\underline{k}_{\{1,\ldots,M_1^1\}}$ can be bounded as
\begin{small}\begin{equation}
A_1<\underline{k}_{\{1,\ldots,M_1^1\}}<\left(\sum_{i=1}^{m_1}k_i^{-1}\right)^{-1},
\end{equation}\end{small}\hspace{-1mm}
where $A_1>0$, and is determined by $\hat{\sigma}_{\{1\}}^2$. The
key observation here is that $\underline{k}_{\{1,\ldots,M_1^1\}}$
is bounded away from $0$ and $+\infty$. The particular value $A_1$
is not important. Using the fact that
$\underline{k}_{\{1,\ldots,M_1^1\}}>0$ and
(\ref{equ:A_K_relation_scalar2}), it is straightforward that
$\alpha_1\sigma_1^2<\alpha_2\sigma_2^2$.

We now discuss the monotonicity properties of
$\underline{k}_{\{1,\ldots,M_1^1\}}$ and $\sigma_2^2$ over
$\sigma_1^2\in(0,\hat{\sigma}_{\{1\}}^2)$. Calculating the
differentiation of (\ref{equ:A_K_relation_scalar}) and
(\ref{equ:A_K_relation_scalar2}) with respect to $\sigma_1^2$
produces
\begin{small}\begin{eqnarray}
&&\left(\frac{\textrm{d}\underline{k}_{\{1,\ldots,M_1^1\}}}{\textrm{d}\sigma_1^2}+1\right)\left(\underline{k}_{\{1,\ldots,M_1^1\}}+\sigma_1^{2}\right)^{-2}=\sum_{i=1}^{m_1}\left(\sigma_1^2+k_{i}\right)^{-2} \label{equ:diff_sigma_1} \\
&&\hspace{-8mm}\textrm{and }-\frac{\textrm{d}\underline{k}_{\{1,\ldots,M_1^1\}}}{\textrm{d}\sigma_1^2}k_{\{1,\ldots,M_1^1\}}^{-2}=-\frac{\alpha_2}{(\alpha_1-\alpha_2)\sigma_1^{4}}+\frac{\alpha_1}{(\alpha_1-\alpha_2)\sigma_2^{4}}\frac{\textrm{d}\sigma_2^2}{\textrm{d}\sigma_1^2}
.\label{equ:diff_sigma_1_2}
\end{eqnarray}\end{small}\hspace{-1mm}
Combining (\ref{equ:A_K_relation_scalar}) and
(\ref{equ:diff_sigma_1}) yields
\begin{small}\begin{equation}
\frac{\textrm{d}\underline{k}_{\{1,\ldots,M_1^1\}}}{\textrm{d}\sigma_1^2}\left(\underline{k}_{\{1,\ldots,M_1^1\}}+\sigma_1^{2}\right)^{-2}=
\sum_{i=1}^{m_1}\left(\sigma_1^2+k_{i}\right)^{-2}-\left(\sum_{i=1}^{m_1}\left(\sigma_1^2+k_{i}\right)^{-1}\right)^2.
\label{equ:diff_sigma_1_3}
\end{equation}\end{small}\hspace{-1mm}
It is easily seen from
(\ref{equ:diff_sigma_1_2})-(\ref{equ:diff_sigma_1_3}) that
$\frac{\textrm{d}\underline{k}_{\{1,\ldots,M_1^1\}}}{\textrm{d}\sigma_1^2}<0$
and $\frac{\textrm{d}\sigma_2^2}{\textrm{d}\sigma_1^2}>0$ when
$\sigma_1^2\in(0,\hat{\sigma}_{\{1\}}^2)$. Thus,
$\underline{k}_{\{1,\ldots,M_1^1\}}$ is monotonically decreasing
over $\sigma_1^2\in(0,\hat{\sigma}_{\{1\}}^2)$, and bounded away
from $0$ and $+\infty$. The parameter $\sigma_2^2$ is
monotonically increasing over
$\sigma_1^2\in(0,\hat{\sigma}_{\{1\}}^2)$, and can take any
positive real value. The parameters $\sigma_1^2$ and $\sigma_2^2$
satisfies the inequality $\alpha_1\sigma_1^2<\alpha_2\sigma_2^2$
for any $\sigma_1^2\in(0,\hat{\sigma}_{\{1\}}^2)$.

\textbf{Case 2 ($j=2,\ldots,J-1$)}: We now study the monotonicity
properties of $\sigma_{j+1}^2$ and $k_{\{1,\ldots,M_1^{j}\}}$ over
$\sigma_1^2$ by assuming that some prior information about
$\sigma_j^2$ and $k_{\{1,\ldots,M_1^{j-1}\}}$ is known.
Specifically, we assume that $\sigma_j^2$ is monotonically
increasing over
$\sigma_1^2\in(0,\hat{\sigma}_{\{1,\ldots,j-1\}}^2)$, and can take
any positive real value. The parameter
$\hat{\sigma}_{\{1,\ldots,j-1\}}^2$ is determined using the
constraint that $\sigma_j^2>\sigma_{j-1}^2$. Also we assume
$\underline{k}_{\{1,\ldots,M_1^{j-1}\}}$ is monotonically
decreasing over
$\sigma_1^2\in(0,\hat{\sigma}_{\{1,\ldots,j-1\}}^2)$, and is
bounded by
$A_{j-1}<\underline{k}_{\{1,\ldots,M_1^{j-1}\}}<\left(\sum_{i=1}^{M_1^{j-1}}k_i^{-1}\right)^{-1}$.

Similarly to the analysis for the case of $j=1$, we first study
the support region of $\sigma_1^2$ such that
$\sigma_{j+1}^2>\sigma_j^2$. Again by combining
(\ref{equ:A_K_relation_scalar})-(\ref{equ:A_K_relation_scalar2}),
there is
\begin{small}\begin{equation}
\frac{\alpha_{j+1}}{\alpha_j\sigma_j^2}+\frac{\alpha_{j+1}-\alpha_j}{\alpha_j(\sigma_{j+1}^2-\sigma_j^2)}=\sum_{j=1}^{m_j}\left(\sigma_j^2+k_{M_1^{j-1}+i}\right)^{-1}
+\left(\sigma_j^2+ k_{M_1^{j-1}}\right)^{-1}.
\end{equation}\end{small}\hspace{-1mm}
We consider the inequality
\begin{small}\begin{eqnarray}
&&\frac{\alpha_{j+1}}{\alpha_j\sigma_j^{2}}>\sum_{i=1}^{m_j}\left(\sigma_j^2+k_{M_1^{j-1}+i}\right)^{-1}+\left(\sigma_j^2+\underline{k}_{\{1,\ldots,M_1^{j-1}\}}
\right)^{-1} \nonumber\\
&&
\stackrel{\sigma_j^2>0}{\Rightarrow}\sum_{i=1}^{m_j}\frac{k_{M_1^{j-1}+i}}{\sigma_j^2+k_{M_1^{j-1}+i}}+\frac{\underline{k}_{\{1,\ldots,M_1^{j-1}\}}}{\sigma_j^2+\underline{k}_{\{1,\ldots,M_1^{j-1}\}}}>m_j+1-\frac{\alpha_{j+1}}{\alpha_j}.
\nonumber\end{eqnarray}\end{small}\hspace{-1mm} Note that
$\sigma_j^2$ and $\underline{k}_{\{1,\ldots,M_1^{j-1}\}}$ are
varying simultaneously along with $\sigma_1^2$. Thus a direct
extension of the analysis for \textbf{Case 1} is unapplicable
here. Fortunately, it is known from the assumption that
$\underline{k}_{\{1,\ldots,M_1^{j-1}\}}$ is bounded away from $0$
and $+\infty$ when
$\sigma_1^2\in(0,\hat{\sigma}_{\{1,\ldots,j-1\}}^2)$. By using the
above observation and the intermediate value theorem as in
\textbf{Case 1}, we conclude that there exists
$0<\hat{\sigma}_{\{1,\ldots,j\}}^2<\hat{\sigma}_{\{1,\ldots,j-1\}}^2$
such that
\begin{small}\begin{eqnarray}
&&0<\sigma_1^2<\hat{\sigma}_{\{1,\ldots,j\}}^2 \Rightarrow \sigma_{j+1}^2>\sigma_j^2  \label{equ:mono_proof3}\\
\textrm{and} && \left\{\begin{array}{l}\sigma_{j+1}^2\rightarrow \sigma_j^2\textrm{ as }\sigma_1^2\rightarrow 0 \\
                        \sigma_{j+1}^2\rightarrow +\infty \textrm{ as } \sigma_1^2\rightarrow \hat{\sigma}_{\{1,\ldots,j\}}^2\end{array}\right.\label{equ:mono_proof4}
\end{eqnarray}\end{small}\hspace{-1mm}
The analysis implies that when the constraint
$\sigma_{j+1}^2>\sigma_j^2$ is imposed, the support region of
$\sigma_1^2$ such that $\sigma_{j}^2>\sigma_{j-1}^2$ becomes
narrow, but still exists. Again by studying
(\ref{equ:A_K_relation_scalar})-(\ref{equ:A_K_relation_scalar2})
under the situation (\ref{equ:mono_proof3}) and
(\ref{equ:mono_proof4}), we arrive at
\begin{small}\begin{equation}
A_j<\underline{k}_{\{1,\ldots,M_1^j\}}<\left(\sum_{i=1}^{M_1^j}k_i^{-1}\right)^{-1},
\end{equation}\end{small}\hspace{-1mm}
where $A_j>0$, and is determined by
$\hat{\sigma}_{\{1,\ldots,j\}}^2$. Using the fact that
$\underline{k}_{\{1,\ldots,M_1^j\}}>0$ and
(\ref{equ:A_K_relation_scalar2}), we obtain
$\alpha_j\sigma_j^2<\alpha_{j+1}\sigma_2^2$ for any $\sigma_1^2\in
(0,\hat{\sigma}_{\{1,\ldots,j\}}^2)$.

Next we consider the monotonicity of $\sigma_{j+1}^2$ and
$\underline{k}_{\{1,\ldots,M_1^{j+1}\}}$ over
$\sigma_1^2\in(0,\hat{\sigma}_{\{1,\ldots,j\}}^2)$. Calculating
the differentiation of
(\ref{equ:A_K_relation_scalar})-(\ref{equ:A_K_relation_scalar2})
w.r.t. $\sigma_1^2$ produces
\begin{small}\begin{eqnarray}
&&\left(\frac{\textrm{d}\underline{k}_{\{1,\ldots,M_1^j\}}}{\textrm{d}\sigma_1^2}+\frac{\textrm{d}\sigma_j^2}{\textrm{d}\sigma_1^2}\right)\left(\underline{k}_{\{1,\ldots,M_1^j\}}+\sigma_j^{2}\right)^{-2}
=\sum_{i=1}^{m_j}\frac{\textrm{d}\sigma_j^2}{\textrm{d}\sigma_1^2}
\left(\sigma_j^2+k_{M_1^{j-1}+i}\right)^{-2} \nonumber\\
&&\hspace{70mm}+\left(\frac{\textrm{d}\sigma_j^2}{\textrm{d}\sigma_1^2}+\frac{\textrm{d}\underline{k}_{\{1,\ldots,M_1^{j-1}\}}}{\textrm{d}\sigma_1^2}\right)\left(\sigma_j^2+\underline{k}_{\{1,\ldots,M_1^{j-1}\}}\right)^{-2}
\label{equ:diff_sigma_j}\\
&&\hspace{-8mm}\textrm{and }-\frac{\textrm{d}\underline{k}_{\{1,\ldots,M_1^j\}}}{\textrm{d}\sigma_1^2}\underline{k}_{\{1,\ldots,M_1^j\}}^{-2}
=-\frac{\textrm{d}\sigma_j^2}{\textrm{d}\sigma_1^2}\frac{\alpha_{j+1}}{(\alpha_j-\alpha_{j+1})\sigma_j^{4}}
+\frac{\textrm{d}\sigma_{j+1}^2}{\textrm{d}\sigma_1^2}\frac{\alpha_j}{(\alpha_j-\alpha_{j+1})\sigma_{j+1}^{4}}.
\label{equ:diff_sigma_j_2}
\end{eqnarray}\end{small}\hspace{-1mm}
Similarly to that of \textbf{Case 1}, By combining
(\ref{equ:A_K_relation_scalar}) and (\ref{equ:diff_sigma_j}), we
have
\begin{small}\begin{eqnarray}
&&\frac{\textrm{d}\underline{k}_{\{1,\ldots,M_1^j\}}}{\textrm{d}\sigma_1^2}\left(\underline{k}_{\{1,\ldots,M_1^j\}}+\sigma_j^{2}\right)^{-2}
=\frac{\textrm{d}\underline{k}_{\{1,\ldots,M_1^{j-1}\}}}{\textrm{d}\sigma_1^2}\left(\sigma_j^2+\underline{k}_{\{1,\ldots,M_1^{j-1}\}}\right)^{-2}
+\frac{\textrm{d}\sigma_j^2}{\textrm{d}\sigma_1^2}\Bigg[\sum_{i=1}^{m_j}
\left(\sigma_j^2+k_{M_1^{j-1}+i}\right)^{-2}\nonumber\\
&&\hspace{0mm}+\left(\sigma_j^2+\underline{k}_{\{1,\ldots,M_1^{j-1}\}}\right)^{-2}-\left(\sum_{i=1}^{m_j}
\left(\sigma_j^2+k_{M_1^{j-1}+i}\right)^{-1}+\left(\sigma_j^2+\underline{k}_{\{1,\ldots,M_1^{j-1}\}}\right)^{-1}\right)^2\Bigg]
\label{equ:diff_sigma_j_3}
\end{eqnarray}\end{small}\hspace{-1mm}
Using the prior information that $\sigma_j^2$ is monotonically
increasing over $\sigma_1^2\in(0,\hat{\sigma}_{\{1,\ldots,j\}}^2)$
and $\underline{k}_{\{1,\ldots,M_1^{j-1}\}}$ is monotonically
decreasing over
$\sigma_1^2\in(0,\hat{\sigma}_{\{1,\ldots,j\}}^2)$, and
(\ref{equ:diff_sigma_j_2})-(\ref{equ:diff_sigma_j_3}), we arrive
at
$\frac{\textrm{d}\underline{k}_{\{1,\ldots,M_1^j\}}}{\textrm{d}\sigma_1^2}<0$
and $\frac{\textrm{d}\sigma_j^2}{\textrm{d}\sigma_1^2}>0$ over
$\sigma_1^2\in(0,\hat{\sigma}_{\{1,\ldots,j\}}^2)$. Thus, the
monotonicity properties of $\underline{k}_{\{1,\ldots,M_1^j\}}$
and $\sigma_j^2$ are proved.

The above induction argument implies that
$\hat{\sigma}_{\{1\}}^2>\hat{\sigma}_{\{1,2\}}^2>\ldots,>\hat{\sigma}_{\{1,\ldots,J-1\}}^2>0$.
Thus when $\sigma_1^2\in(0,\hat{\sigma}_{\{1,\ldots,J-1\}}^2)$, it
is guaranteed that (\ref{equ:A_inequality_scalar}) holds. The
inequality (\ref{equ:A_inequality_scalar}) plays an important role
in showing that the Gaussian test channel achieves the optimal sum
rate.

\textbf{Case 3 (monotonicity of $\underline{k}_w$ (or
$\underline{k}_{\{1,\ldots,L\}}$))}: We list the prior conditions
for the derivation. From the analysis in \textbf{Case 1} and
\textbf{2}, it is clear that
$\underline{k}_{\{1,\ldots,M_1^{J-1}\}}$ is monotonically
decreasing over
$\sigma_1^2\in(0,\hat{\sigma}_{\{1,\ldots,J-1\}}^2)$, and bounded
away from $0$ and $+\infty$, i.e.
$A_{J-1}<\underline{k}_{\{1,\ldots,M_1^{J-1}\}}<\left(\sum_{i=1}^{M_1^{J-1}}k_i^{-1}\right)^{-1}$.
Also $\sigma_J^2$ is monotonically increasing over
$\sigma_1^2\in\left(0,\hat{\sigma}_{\{1,\ldots,J-1\}}^2\right)$,
and can take any positive real value.

We work with (\ref{equ:A_K_relation_scalar}). In order that
$\underline{k}_0$ is positive, we consider the inequality
\begin{small}\begin{eqnarray}
&&\sigma_J^{-2}>\left(\sigma_J^2+\underline{k}_{\{1,\ldots,M_1^{J-1}\}}\right)^{-1}+\sum_{i=1}^{m_J}\left(\sigma_J^2+k_{M_1^{J-1}+i}\right)^{-1}
\nonumber \\
&& \stackrel{\sigma_J^2>0}{\Rightarrow}\sum_{i=1}^{m_J}\frac{k_{M_1^{J-1}+i}}{\sigma_J^2+k_{M_1^{J-1}+i}}+\frac{\underline{k}_{\{1,\ldots,M_1^{J-1}\}}}{\sigma_J^2+\underline{k}_{\{1,\ldots,M_1^{J-1}\}}}
>m_J.
\end{eqnarray}\end{small}\hspace{-1mm}
Again using the fact that $\underline{k}_{\{1,\ldots,M_1^{J-1}\}}$
is bounded away from 0 and $+\infty$, and the intermediate value
theorem, we conclude that there exists
$0<\ddot{\sigma}_1^2<\hat{\sigma}_{\{1,\ldots,J-1\}}^2$ such that
\begin{small}\begin{eqnarray}
&& 0<\sigma_1^2<\ddot{\sigma}_1^2\Rightarrow \underline{k}_w>0 \\
\textrm{and }&& \left\{\begin{array}{l}
\underline{k}_w\rightarrow \left(\sum_{i=1}^{L}k_i^{-1}\right)^{-1}\textrm{ as }\sigma_1^2\rightarrow 0\\
\underline{k}_w\rightarrow 0\textrm{ as }\sigma_1^2\rightarrow \ddot{\sigma}_1^2
\end{array}\right. .
\end{eqnarray}\end{small}\hspace{-1mm}
To show that $\underline{k}_w$ is monotonically increasing over
$\sigma_1^2\in(0,\ddot{\sigma}_1^2)$, one can take the
differentiation of (\ref{equ:A_K_relation_scalar}) w.r.t.
$\sigma_1^2$ and use the monotonicity properties of $\sigma_J^2$
and $\underline{k}_{\{1,\ldots,M_1^{J-1}\}}$. The argument is
straightforward. The proof is complete.

\section{Proof of Lemma \ref{lemma:Kw_positive}}
\label{lemma:Kw_positive_proof} To prove that the corresponding
$\boldsymbol{K}_w$ is positive definite, it is sufficient to show
that $|\boldsymbol{K}_{\{1,\ldots,j\}}|> 0$, $\forall$
$j=1,\ldots,L$. Before presenting the proof, we first show that
$\underline{\boldsymbol{K}}_{\{1,\ldots,M_1^j\}}$, $j=1,\ldots,J$
are positive definite. Using the fact that
$\boldsymbol{0}\prec\alpha_1\boldsymbol{A}_1\prec
\alpha_2\boldsymbol{A}_2(\boldsymbol{A}_1)\prec\ldots
\prec\alpha_J\boldsymbol{A}_J(\boldsymbol{A}_1)$ and
(\ref{equ:proportionality}), we arrive at
\begin{small}\begin{equation}
\underline{\boldsymbol{K}}_w\prec \underline{\boldsymbol{K}}_{\{1,\ldots,M_1^{J-1}\}}\prec\ldots
\prec  \underline{\boldsymbol{K}}_{\{1,\ldots,M_1^{1}\}}.
\label{equ:K_ordering}
\end{equation}\end{small}\hspace{-1mm}

We now use induction argument to prove the lemma. Assume
$\boldsymbol{K}_{\{1,\ldots,M_1^{j-1}\}}\succ\boldsymbol{0}$, we
show that the determinant of
$\boldsymbol{K}_{\{1,\ldots,M_1^{j-1}+i\}}$ is positive for any
$i=1,\ldots,m_j$. For the special case that $j=1$, we make no
assumption.  The argument is provided as follows:
\begin{small}\begin{eqnarray}
&& \hspace{-6mm}\Big|\boldsymbol{K}_{\{1,\ldots,M_1^{j-1}+i\}}\Big| \nonumber \\
&=&\hspace{-2mm}\Big|\textrm{diag}\Big(\boldsymbol{K}_{\{1,\ldots,M_1^{j-1}\}}+\boldsymbol{H}\otimes \boldsymbol{A}_j,
\boldsymbol{K}_{M_1^{j-1}+1}+\boldsymbol{A}_j,\ldots,\boldsymbol{K}_{M_1^{j-1}+i}+\boldsymbol{A}_j\Big)
-(\boldsymbol{I}_N,\ldots,\boldsymbol{I}_N)^t\boldsymbol{A}_j(\boldsymbol{I}_N,\ldots,\boldsymbol{I}_N)\Big| \nonumber\\
&=&\hspace{-2mm}\Big|\boldsymbol{I}_{N(M_1^{j-1}+i)}-
(\boldsymbol{I}_N,\ldots,\boldsymbol{I}_N)^t\boldsymbol{A}_j(\boldsymbol{I}_N,\ldots,\boldsymbol{I}_N)
\cdot\textrm{diag}\Big(\left(\boldsymbol{K}_{\{1,\ldots,M_1^{j-1}\}}+\boldsymbol{H}\otimes \boldsymbol{A}_j\right)^{-1},
\nonumber\\
&&
\left(\boldsymbol{K}_{M_1^{j-1}+1}+\boldsymbol{A}_j\right)^{-1},\ldots,\left(\boldsymbol{K}_{M_1^{j-1}+i}+\boldsymbol{A}_j\right)^{-1}\Big)\Big|\cdot\Big|\boldsymbol{K}_{\{1,\ldots,M_1^{j-1}\}}+\boldsymbol{H}\otimes
\boldsymbol{A}_j\Big|
\cdot\prod_{k=1}^{i}\Big|\boldsymbol{K}_{M_1^{j-1}+k}+\boldsymbol{A}_j\Big| \nonumber\\
&\stackrel{(a)}{=}& \hspace{-2mm}\Big|\boldsymbol{A}_j\Big|\cdot\Big|\boldsymbol{A}_j^{-1}-\Big(\sum_{k=1}^i\left(\boldsymbol{K}_{M_1^{j-1}+k}+\boldsymbol{A}_j\right)^{-1}
+\left(\underline{\boldsymbol{K}}_{\{1,\ldots,M_1^{j-1}\}}+\boldsymbol{A}_j\right)^{-1}\Big)\Big|\cdot\Big|\boldsymbol{K}_{\{1,\ldots,M_1^{j-1}\}}+\boldsymbol{H}\otimes \boldsymbol{A}_j\Big|
\nonumber \\
&&\cdot\prod_{k=1}^{i}\Big|\boldsymbol{K}_{M_1^{j-1}+k}+\boldsymbol{A}_j\Big|,
\label{equ:det_positive}
\end{eqnarray}\end{small}\hspace{-1mm}
where (a) follows from (\ref{equ:proof_A_K_relation_3}) and the
determinant identity $|\boldsymbol{I}_{m}-\boldsymbol{A}_{m\times
n}\boldsymbol{B}_{n\times
m}|=|\boldsymbol{I}_n-\boldsymbol{B}_{n\times
m}\boldsymbol{A}_{m\times n}|$. Since
$\boldsymbol{K}_{\{1,\ldots,M_1^{j-1}\}}$ is assumed to be
positive definite, it is straightforward that
$\left|\boldsymbol{K}_{\{1,\ldots,M_1^{j-1}\}}+\boldsymbol{H}\otimes
\boldsymbol{A}_j\right|>0$. From the condition of the lemma, it is
also known that $\big|\boldsymbol{A}_j\big|>0$ and
$\big|\boldsymbol{K}_{M_1^{j-1}+k}+\boldsymbol{A}_j\big|>0$,
$k=1,\ldots,m_j$. Now we prove that the remaining quantity in
(\ref{equ:det_positive}) is positive. From
(\ref{equ:A_K_relation}), there is
\begin{small}\begin{eqnarray}
&&\hspace{-7mm}\left(\underline{\boldsymbol{K}}_{\{1,\ldots,M_1^{j}\}}+\boldsymbol{A}_j\right)^{-1}=\sum_{k=1}^{m_j}\left(\boldsymbol{K}_{M_1^{j-1}+k}+\boldsymbol{A}_j\right)^{-1}
+\left(\underline{\boldsymbol{K}}_{\{1,\ldots,M_1^{j-1}\}}+\boldsymbol{A}_j\right)^{-1}\nonumber \\
&\Rightarrow & \left(\underline{\boldsymbol{K}}_{\{1,\ldots,M_1^{j}\}}+\boldsymbol{A}_j\right)^{-1}\succeq \sum_{k=1}^i\left(\boldsymbol{K}_{M_1^{j-1}+k}+\boldsymbol{A}_j\right)^{-1}
+\left(\underline{\boldsymbol{K}}_{\{1,\ldots,M_1^{j-1}\}}+\boldsymbol{A}_j\right)^{-1}\nonumber \\
&\stackrel{(a)}{\Rightarrow} & \boldsymbol{A}_j^{-1}\succ \sum_{k=1}^i\left(\boldsymbol{K}_{M_1^{j-1}+k}+\boldsymbol{A}_j\right)^{-1}
+\left(\underline{\boldsymbol{K}}_{\{1,\ldots,M_1^{j-1}\}}+\boldsymbol{A}_j\right)^{-1},\nonumber
\end{eqnarray}\end{small}\hspace{-1mm}
where $(a)$ follows from (\ref{equ:K_ordering}) and Lemma
\ref{lemma:ordering_relation}. It is immediate that
$\big|\boldsymbol{A}_j^{-1}-
\sum_{k=1}^i\left(\boldsymbol{K}_{M_1^{j-1}+k}+\boldsymbol{A}_j\right)^{-1}
+\left(\underline{\boldsymbol{K}}_{\{1,\ldots,M_1^{j-1}\}}+\boldsymbol{A}_j\right)^{-1}\big|>0$.
Thus we have proven that the determinant of
$\boldsymbol{K}_{\{1,\ldots,M_1^{j-1}+i\}}$ is positive for any
$i=1,\ldots,m_j$, $j=1,\ldots,J$. We conclude that the considered
$\boldsymbol{K}_w$ is positive definite. The proof is complete.

\ifCLASSOPTIONcaptionsoff
  \newpage
\fi



\bibliographystyle{IEEEtran}
%

%




\end{document}